\documentclass[preprint,showpacs,preprintnumbers,amsmath,xamssymb,aps,pop]{revtex4-1}

\usepackage{amssymb,amsmath,amsfonts,amstext,graphics,graphicx, subfigure}

\newenvironment{proof}[1][Proof]{\noindent\textbf{#1.} }{\ \rule{0.5em}{0.5em}}

\def\XXint#1#2#3{{\setbox0=\hbox{$#1{#2#3}{\int}$ }
\vcenter{\hbox{$#2#3$ }}\kern-.5\wd0}}

\def\calc{\mathcal{C}}

\def\cale{\mathcal{E}}
\def\calf{\mathcal{F}}
\def\calg{\mathcal{G}}
\def\calh{\mathcal{H}}
\def\cali{\mathcal{I}}
\def\calj{\mathcal{J}}
\def\calk{\mathcal{K}}

\def\calp{\mathcal{P}}

\def\R{\mathbb{R}}

\def\E{{\rm I\kern-.1567em E}}
\def\A{{\rm I\kern-.1567em A}}
\def\P{{\rm I\kern-.1567em P}}
\def\V{{\rm I\kern-.1567em V}}

\def\bq{\begin{equation}}
\def\eq{\end{equation}}
\def\bqy{\begin{eqnarray}}
\def\eqy{\end{eqnarray}}

\def\tensep{\underline{\underline{\varep}}}
\def\tensepl{\underline{\underline{\ep}}}
\def\tensmu{\underline{\underline{\mu}}}
\def\tensk{\underline{\underline{k}}}
\def\tensi{\underline{\underline{I}}}
\def\tensche{\underline{\underline{\chi_{e}}}}

\def\al{\alpha}
\def\be{\beta}

\def\de{\delta}

\def\ep{\epsilon}

\def\ga{\gamma}

\def\ph{\phi}

\def\varep{\varepsilon}

\def\bfb{\mathbf{B}}
\def\bfd{\mathbf{D}}
\def\bfe{\mathbf{E}}
\def\bfa{\mathbf{A}}
\def\bfj{\mathbf{J}}
\def\bfP{\mathbf{P}}
\def\bfm{\mathbf{M}}
\def\bfh{\mathbf{H}}

\def\bfx{\mathbf{x}}
\def\bfv{\mathbf{v}}
\def\bfp{\mathbf{p}}

\def\ub{\mathbf{b}}
\def\bfs{\mathbf{s}}

\def\p{\partial}

  \def\Brac#1#2{\{#1,#2\}}
  \def\brac#1#2{[#1,#2]}
  \def\nest#1#2#3{\{\{#1,#2\},#3\}}
  
\def\ncr{\nonumber\\}

\def\deq{\ \dot=\ }

\begin{document}

\title{A general theory for gauge-free lifting}
\author{P.~J.~Morrison}
 \email{morrison@physics.utexas.edu}
\affiliation{Department of Physics and Institute for Fusion Studies, 
The University of Texas at Austin, Austin, TX, 78712, USA}
\date{\today}

\begin{abstract}

A theory for lifting equations of motion for charged particle dynamics, subject to given electromagnetic like forces,  up to a gauge-free system of coupled Hamiltonian Vlasov-Maxwell like equations is given.  The theory provides very general expressions for the polarization and magnetization vector fields in terms of the particle dynamics description of matter.  Thus,  as is common in plasma physics, the particle dynamics replaces conventional constitutive relations for matter.  Several examples are considered including the usual Vlasov-Maxwell theory, a guiding center kinetic theory, Vlasov-Maxwell theory with the inclusion of spin, and a Vlasov-Maxwell theory with the inclusion of Dirac's magnetic monopoles.   All are shown to be Hamiltonian field theories and the Jacobi identity is  proven directly. 

\bigskip

\end{abstract}

\maketitle


\section{Introduction}
\label{intro}

In conventional treatments of electricity and magnetism  phenomenological susceptibilities are introduced to describe material  media.  Concomitant with the introduction of these susceptibilities is the idea that charge  can be  separated into bound and free components, current can be similarly decomposed, and based on these separations expressions for the polarization and magnetization of the medium are  obtained.   However, it is well-known to plasma physicists that such a simple characterization is not possible for plasmas, where particle orbits may transition from trapped to passing and, indeed,  may exhibit complicated behavior that can only be described by the self-consistent treatment of the dynamics of both the particles and the fields.  Because of these complications,  tractable and reliable  expressions for the polarization and magnetization are not so forthcoming, particularly when approximations are made and/or additional physics is added. 

The purpose of the present paper is to construct a general theory for the coupling of charge carrying  particle dynamics, entities possibly with internal degrees of freedom described by a kinetic theory, coupled to electromagnetic-like field theories.  A theory that is  gauge-free and ultimately expressible without the introduction of  vector and scalar potentials is constructed.  Like Maxwell's equations, the Vlasov-Maxwell  equations, and virtually every important system in physics, the  theory will have  Hamiltonian form.  This Hamiltonian form will be noncanonical, following the program initiated  in Refs.~\cite{morgreene,morrison80}. 

The construction begins in Sec.~\ref{genkentheory} with a set of ordinary differential equations that describe the particle dynamics.   This set of equations, which is the basic model of the matter under consideration,  is assumed to have a very general Hamiltonian form, possibly with an unconventional phase space  and with a Hamiltonian  that depends  on  specified electromagnetic fields including the  field variables $\mathbf{E}(\mathbf{x},t)$ and $\mathbf{B}(\mathbf{x},t)$, and  possibly all their derivatives.  The problem then is to {\it lift} this finite-dimensional dynamical system that describes the matter to a gauge-free field theory with  a kinetic component that is  of Vlasov type coupled to an electromagnetic component  of Maxwell type.    The difficulty with this lifting program lies in the coupling of the two components of the field theory.   It is shown in Sec.~\ref{ncbkt} that the construction given naturally results in a  field theory that is also Hamiltonian.  This assures  that there is a consistency to the coupling.  Because the Hamiltonian theory requires variational calculus, it is most convenient to discuss constitutive relations resulting from the matter system in this section as well.  In Sec.~\ref{examples}  several examples are presented, beginning with the usual Vlasov-Maxwell system, followed by  a general guiding center kinetic theory, a theory that includes spin  and, to show the generality of our construction,  a theory with monopole charge where the Maxwell field is modified.  Gaussian units are used for all examples. Section \ref{conclusions}   contains concluding remarks.    In Appendix \ref{Ajacobi} of the paper there are  several subsections with direct proofs of the Jacobi identity for  Poisson brackets of the noncanonical Hamiltonian field theories.   The first one  describes  an old calculation  of the author that has not heretofore appeared in print, a calculation that contains several useful techniques.  The other subsections contain analyses of the  other  brackets of  the  examples of Sec.~\ref{examples}.


\section{A General Electromagnetic Kinetic Theory via Lifting}
\label{genkentheory}


Consider a general dynamical system with an  $n$-dimensional  phase space with coordinates $z=(z^1,z^2,\dots, z^n)$ and evolution determined by a Poisson bracket and Hamiltonian $\cale$ as follows:
\bq
\dot z^a=[z^a,\cale]=J^{ab}\frac{\p \cale}{\p z^b}\,,\qquad a,b=1,2,\dots, n
\label{ndyn}
\eq
where the Poisson bracket on phase space functions $g$ and $h$ is defined by
\bq
[g,h]=\frac{\p g}{\p z^a}J^{ab}\frac{\p h}{\p z^b}
\label{npb}
\eq
and repeated indices are to be summed.  The only requirements placed on the cosymplectic tensor $J$ is that it endow the Poisson bracket with the properties of antisymmetry,  $[g,h]=-[h,g]$,  and the Jacobi identity, 
$[g,[h,k]]+ [h,[k,g]] +  [k,[g,h]]=0$ for all functions $g,h,k$ (see e.g.\ \cite{morrison82,morrison98}).  In the context of geometrical  mechanics this is referred to as a flow on a Poisson manifold, but this formalism and its language will not be used  here.  Rather, the physics of matter described by this finite-dimensional dynamical system, as embodied in the Hamiltonian function $\cale$,  is emphasized.  The description of the matter  in the  formalism of this paper is  contained in this function $\cale$ and its associated Poisson bracket (\ref{npb}).

Particle orbits in given $\mathbf{E}(\mathbf{x},t)$ and $\mathbf{B}(\mathbf{x},t)$ fields are usually described in terms of the  electromagnetic  potentials, $\phi(\mathbf{x},t)$ and $\mathbf{A}(\mathbf{x},t)$, where 
\bq
 \mathbf{E}= -\nabla \phi -\frac1{c}\frac{\p  \mathbf{A}}{\p t}\quad {\rm and}\quad 
  \mathbf{B}=\nabla\times  \mathbf{A}\,.
\eq
Following this usual procedure,  the  Hamiltonian $\cale$ of the general  system  of (\ref{ndyn})  will be restricted for the purposes of the present lift theory to have the following form:  
\bq
\cale=\bar\calk \left(\mathbf{p}- {e\mathbf{A}}/{c},w, \mathbf{E},\mathbf{B}, \nabla \bfe,\nabla\bfb, \dots\right) + e\phi\,, 
\label{gHam}
\eq
where $\bar\calk$ is an arbitrary function of its arguments.  This form was proposed in the context of the variational theory of Refs.~\cite{P84,PM85,PM91}.  Since the aim is to generalize usual charged particle dynamics,  the parameter $e$ that denotes charge is included and $c$ is the speed of light as usual.    This particular form  assures electromagnetic gauge invariance.  Here,  the phase space has been split into two parts, 
$z_p=(\bfx,\bfp, w^1,w^2,\dots, w^d)$,  where  the coordinates $(\bfx,\bfp)$ are the usual canonical six-dimensional phase space coordinates and the coordinates $w=(w^1,w^2,\dots, w^d)$ describe additional degrees of freedom, such as might occur in the classical description of a molecule  with rotational or vibrational degrees of freedom.     These additional coordinates will be referred to as {\it internal}   degrees of freedom.    An important, but not the only, Poisson bracket is given by 
\bqy
 [f,g]&=\ &  [f,g]_{\bfp} + [f,g]_{w}
 \nonumber\\
   &=:&\nabla f \cdot\frac{\p g}{\p \mathbf{p}}-  \nabla g \cdot\frac{\p f}{\p \mathbf{p}} 
 + \frac{\p f}{\p w^{\alpha}}J_w^{\al \be} \frac{\p g}{\p w^{\be}}\,, \qquad \al,\be=1,2,\dots, d\,,
 \label{splitpbkt}
 \eqy
which is the sum of canonical and internal pieces, and is assumed to satisfy the Jacobi identity.  Such a bracket and Hamiltonian generate  dynamics of the form of (\ref{ndyn}). 

Several comments on the Hamiltonian form of (\ref{gHam}) and (\ref{splitpbkt}) are in order.  Note that $\cale$ may depend explicitly on the fields, and all of their derivatives, and the same may be true for the tensor $J$ of the Poisson bracket provided the  Jacobi identity is satisfied for any choice of these fields.   The Poisson bracket may also have explicit $w$ dependence, but  no direct coupling of the internal  degrees of freedom to the fields has been made explicit.  In general, such coupling  would need to be consistent with the symmetries of interest for these variables.  For example, if $w$ were a spin variable,  say $\bfs$, then  $\cale$ would depend  on $\bfs\cdot \bfb$.  This case is treated as an example in Sec.~\ref{spin}.   Also,  observe that  all of the dependence on the spatial variable  $\mathbf{x}$ occurs through the fields.  The  omission of such explicit  dependence in $\cale$  is appropriate for media where  spatial homogeneity is broken only by the presence of the fields;  however,  the $\mathbf{x}$ dependence  could be added for further generalization.  

Alternatively,  a manifestly gauge invariant form is obtained in terms of the coordinates
$z_v=(\bfx,\bfv, w^1,w^2,\dots, w^d)$, 
where $(\bfx,\bfv)$ denotes  the usual  six-dimensional velocity phase space coordinates, with   
\bq
\mathbf{p}:=m \bfv + \frac{e}{c}\mathbf{A} \,, 
 \label{v-p}
\eq
and $m$ denoting the mass of the charged particle.  In terms of these variables (\ref{ndyn}) takes the form
\bq
\dot z_v=[z_v,\calk] + \frac{e}{m} \cali_d\cdot\mathbf{E}\,,
\label{spliteom}
\eq
where the bracket of (\ref{splitpbkt}) becomes the Littlejohn  \cite{littlejohn} Poisson bracket  in (\ref{spliteom}),
\bq
\left[g,h\right]_{L}= [g,h]_{\bfv} + [g,h]_{\bfb}\,,
\label{nc1}
\eq
 where
\bq
[g,h]_{\bfv}=\frac1{m}\left(\nabla g\cdot \frac{\p h}{\p \bfv}   - \nabla h \cdot\frac{\p g}{\p \bfv}\right) 
\quad {\rm and} \quad  
[g,h]_{\bfb} = \frac{e}{m^2c}\bfb\cdot\left(\frac{\p g}{\p \bfv}\times\frac{\p h}{\p \bfv}\right)\,, 
\label{nc2}
\eq
 $\cali_d$ is a $(6+ d)\times 3$ matrix used to embed $\bfe$ into the force law, and 
\bq
\calk \left(\mathbf{v},w; \mathbf{E},\mathbf{B}, \nabla \bfe,\nabla\bfb, \dots\right) 
=\bar\calk \left(\mathbf{p}- {e\mathbf{A}}/{c},w, \mathbf{E},\mathbf{B}, \nabla \bfe,\nabla\bfb, \dots\right)\,.
\label{calk}
\eq
Thus the electric field appears as an external force  in addition to any dependence on it that may come  through  the function $\calk$, and the electromagnetic potentials  no longer appear in the dynamics.  Note,  in general  $\bfe$   cannot be written as a gradient in order to combine it with the first term of (\ref{spliteom}).    The dynamics of (\ref{spliteom}) with arbitrary Poisson bracket in terms of $z_v$,  possibly depending explicitly on $z_v$, $\bfe$, and $\bfb$, can be taken as the starting point for lifting to a kinetic theory. 

Usual  Lorentzian dynamics is given by $\calk=m|\mathbf{v}|^2/2$:  when $\calk$ is  written in terms of $\bfp$, the bracket of (\ref{splitpbkt}) with $\cale$ yields the equations of motion for a  particle of charge $e$ and mass $m$ subject to given electric and magnetic fields.  Alternatively, the same equations are given from  (\ref{spliteom}) with $[\ ,\ ]_L$.  


\medskip

Now,  the finite degree-of-freedom  system of  (\ref{spliteom})  will be lifted.  The first step  is to lift the particle dynamics  to a kinetic theory for determining  a phase space density $f(z,t)=f(\mathbf{x},\mathbf{v},w,t)$.  This  is easily achieved by the standard Liouville form 
 \bq
 \frac{\p  f}{\p t} + \left[f,\calk\right] + \frac{e}{m}\, \bfe\cdot \frac{\p f}{\p \bfv} =0\,,
 \label{genken}
 \eq
where the generalization to multiple species is straightforward.  Clearly the characteristic equations  of (\ref{genken}) correspond to the  finite-dimensional matter model  of (\ref{spliteom}).   

The second part of lifting is to  describe the coupling to Maxwell's equations.   This coupling is effected by introducing the energy functional
\bq
K[\bfe,\bfb,f]:=\int\! \!d\bfx d\bfv dw\, \calk\,  f\,,
\eq
whence the following expressions for the charge and current densities are obtained:
 \bqy
 \rho(\mathbf{x},t)&=&e\int \!\!d\bfv dw\,   f  - \nabla\cdot  \frac{\de K}{\de \bfe}
 \label{density}
 \\  
 \mathbf{J}(\mathbf{x},t)&=&e\int\! \!d\bfv dw\,\frac{\p \calk}{\p \mathbf{v}}\,   f 
+ \frac{\p }{\p t}   \frac{\de K}{\de \bfe}
+ c\,  \nabla\times  \frac{\de K}{\de \bfb}\,.
\label{current}
\eqy
Inserting these expressions for the sources into the usual form of Maxwell's equations completes the lift.  

From (\ref{density}) and (\ref{current}) it  is evident that   the polarization, $\mathbf{P}$, and magnetization, $\mathbf{M}$,  can be identified as 
\bq
\mathbf{P}(\mathbf{x},t)= -  \frac{\de K}{\de \bfe}
 \quad{\rm and}\quad
 \mathbf{M}(\mathbf{x},t)= -   \frac{\de K}{\de \bfb}\,,
 \label{PM}
\eq
which is consistent with the usual definitions of bound charge density, polarization current, and magnetization current,
\bq
\rho_b=-\nabla\cdot \bfP\,, \quad\mathbf{J}_p=  \frac{\p \mathbf{P}}{\p t}\,,\quad 
{\rm and}\quad 
\mathbf{J}_m=c\nabla\times \mathbf{M}\,, 
\eq
respectively. 
Although the manner of lifting embodied in  (\ref{genken}), (\ref{density}),  and (\ref{current})  is straightforward, because of the functional derivatives in (\ref{PM}) the dependencies  of $\mathbf P$ and $\mathbf M$ on the fields $\mathbf{E}$ and  $\mathbf{B}$ may be very complicated and contain high order spatial derivatives.  

In Sec.~\ref{ncbkt} it is  shown that this manner of lifting results in a Hamiltonian field description of the coupled system.  It should be emphasized  that this construction does not require the explicit introduction of the vector and scalar potentials.  However, from the Hamiltonian form using $\cale$ it is clear that it subsumes the description using  $\phi(\mathbf{x},t)$ and $\mathbf{A}(\mathbf{x},t))$.  To see the explicit form, define  the momentum phase space density by  $\bar{f}(\mathbf{x},\mathbf{p},w, t)= f(\mathbf{x},\mathbf{v},w, t)$,  which gives under the change $\bfv \leftrightarrow \bfp$ the  governing kinetic equation 
 \bq
 \frac{\p \bar f}{\p t} + \left[\bar f,\cale\right]=0\,,
 \label{bgenken}
 \eq
with $ [g,h]= [f,g]_{\bfp} + [f,g]_{w}$. 
 The coupling to Maxwell's equation is essentially unchanged:  
\bqy
 \rho(\mathbf{x},t)&=&e\int \!\!d\bfp dw\, \bar f  - \nabla\cdot  \frac{\de K}{\de \bfe}
 \\  
 \mathbf{J}(\mathbf{x},t)&=&e\int\! \!d\bfp dw\,\frac{\p \calk}{\p \mathbf{p}}\, \bar f 
+ \frac{\p }{\p t}   \frac{\de K}{\de \bfe}
+ c \nabla\times  \frac{\de K}{\de \bfb}\,, 
\label{newJ}
\eqy
as are the expressions for  $\mathbf{P}$ and  $\mathbf{M}$. 
Using the chain rule expressions, 
 \bq
  \frac{\p  \bar  f}{\p  \mathbf{p}}= \frac1{m}\frac{\p   f}{\p  \mathbf{v}}   \,,\quad 
  \nabla {\bar  f}=  \nabla  {f} - \frac{e}{mc}\, \frac{\p {f}}{\p  \bfv}\cdot \nabla \mathbf{A}\,, 
\quad {\rm and} \quad
 \frac{\p \bar  f}{\p t}=  \frac{\p  {f}}{\p t} -\frac{e}{mc} \, \frac{\p  f}{\p  \mathbf{v}}\cdot \frac{\p \mathbf{A}}{\p t}\,, 
  \eq
it is not difficult to show that $\left[g,h \right]_{\bfp}$ transforms to 
$\left[g,h \right]_{L}$  and (\ref{bgenken}) transforms into  (\ref{genken}).

  
\section{Hamiltonian Form and  Constitutive Relations}
\label{ncbkt}

Based on past experience, viz.\ Vlasov-Maxwell and guiding center kinetic theories, a natural choice for the  Hamiltonian functional is the following: 
\bqy
H[f, \mathbf{E}, \mathbf{B}]&=&  K - \int\!\!d\bfx \, \bfe \cdot \frac{\de K}{\de \bfe} + {\frac1{8\pi}}\int\!\!d\bfx\left(|\bfe|^2 + |\bfb|^2\right)
\nonumber\\
&=&  K + \int\! \!d{\bfx}\,  \bfe\cdot \bfP  + {\frac1{8\pi}}\int\!\!d\bfx\left(|\bfe|^2 + |\bfb|^2\right)
\label{ham}
\eqy
where $\bfP$, as given by (\ref{PM}),  is used as a shorthand in the second line, which one could rewrite in terms of  $\bfd:= \bfe +4\pi\bfP$.  The Hamiltonian of (\ref{ham}) is a generalization of the energy  component of the energy-momentum tensor first derived by variational methods in \cite{P84,PM85,PM91}.   It is straightforward to  verify directly that   (\ref{ham})  is conserved by the combined field theory, Maxwell's equations with the sources (\ref{density}) and (\ref{current}) coupled to  the kinetic theory of  (\ref{genken}).

One might think that the  $|\bfb|^2$ term of (\ref{ham}) should be replaced by $ \mathbf{B}\cdot  \mathbf{H}$, where $\bfh=\bfb-4\pi\bfm$, but this is incorrect.    All polarization and magnetization effects are  modeled here by the  terms involving $\calk$, i.e., they are a consequence of  the particle dynamics.  Rather than relating $ \mathbf{E}$ and $\mathbf{B}$ to  $\mathbf{D}$ and $\mathbf{H}$ by constitutive relations, the  particle dynamics, extended and other,  describes the physics that is often  approximated by simplistic constitutive relations.  For example, the difference between $|\bfe|^2$ and $ \mathbf{E}\cdot  \mathbf{D}$ arises from  the $\calk$ term that contains the matter dynamical information. 

If only $(f, \mathbf{E}, \mathbf{B})$ are used as  dynamical variables, there is a difficulty in obtaining a Poisson bracket description for the field theory.  The problem is readily encountered when one attempts to include polarization effects, because the polarization current has a time derivative and Poisson bracket expressions such as $\{\bfe, H\}$  do not produce terms with time derivatives of the dynamical variables; i.e., in Hamiltonian theories all time derivatives are on the left hand side, so to speak. Consequently a term of the form $\p \bfP/\p t$ cannot appear.  However, there is a  way to circumvent this problem, a problem that does not occur in action principle formulations, such as those of \cite{P84,PM85,PM91}.  

Functional differentiation of  (\ref{ham}) gives 
\bqy 
\frac{\de H}{\de \mathbf{E}}&=&-  \left(\frac{\de^2 K}{\de \bfe \de \bfe}\right)^{\dagger} \!\!\cdot \bfe +  \frac{\mathbf{E}}{4\pi}\,,
 \label{DHE}\\
\frac{\de H}{\de \mathbf{B}}&=&\frac{\de K}{\de \bfb} - \left(\frac{\de^2 K}{\de \bfb \de \bfe}\right)^{\dagger}\!\!\cdot \bfe  +\frac{\bfb}{4\pi}\,,
 \label{DHB}
\eqy
where ${\de^2 K}/{\de \bfe \de \bfe}$  and ${\de^2 K}/{\de \bfe \de \bfb}$ are second functional derivative operators that satisfy
\bq
\left(\frac{\de^2 K}{\de \bfe \de \bfe}\right)^{\dagger}=\frac{\de^2 K}{\de \bfe \de \bfe}
\quad {\rm and}\quad
\left(\frac{\de^2 K}{\de \bfb \de \bfe}\right)^{\dagger}=\frac{\de^2 K}{\de \bfe \de \bfb}\,.
\eq
(See e.g.\ \cite{morrison82,morrison98,morrison05} for a review of functional differentiation.)   Expressions (\ref{DHE}) and  (\ref{DHB}) reveal how polarization and magnetization effects are embodied in $\calk$.   Since the functional derivatives above have, in a sense, `dressed' $\mathbf{E}$ and $\mathbf{B}$, existing Hamiltonian structures will not be adequate.  It is clear that without some modification one cannot obtain the polarization current.

If the theory were expressed in terms of $\bfd$,  the following bracket on functionals  $\bar{F}[\bfd,\bfb,f]$ could give the correct temporal evolution of $\bfd$,  provided $\de \bar{H}/\de \bfb=\bfh/4\pi$:
\bqy
\{\bar{F},\bar{G}\}&=&\int\!\!d\bfx d\bfv  dw\, f \left(  [\bar F_f,\bar{G}_f]_{\bfv} + [\bar F_f,\bar{G}_f]_{w}\right)
\label{pjm}
\\
&{\ }&  + \int\!\!d\bfx d\bfv  dw\, f \,  [\bar F_f,\bar{G}_f]_{\bfb}
\label{MW}
\\
 &{\ }& \hspace{ .5 in} 
+ \frac{4\pi e}{m}\int\!\!d\bfx d\bfv  dw\, f \left( \bar G_{\bfd}\cdot\p_{\bfv} \bar F_f  - \bar F_{\bfd}\cdot\p_{\bfv} \bar G_f\right)
\label{coupling}
\\
&{\  }& \hspace{1.0 in} + 4\pi c \int\!\!d \bfx \, \big(\bar F_{\bfd}\cdot \nabla\times  \bar G_{\bfb}
- \bar G_{\bfd}\cdot \nabla\times  \bar F_{\bfb}\big)\,, 
\label{dbkt}
\eqy
with $[f,g]_{\bfv}$  and  $[f,g]_{\bfb}$  given by  (\ref{nc2}), $\p_{\bfv} g=\p g/\p \bfv$, 
and $\bar{F}_f:=\de \bar F/\de f$, $\bar{F}_{\bfd}:=\de \bar F/\de \bfd$,  etc.  The  Born-Infeld  term  of (\ref{dbkt}) is motivated by their original  theory \cite{bornin} that was also written in terms of $\bfd$ and $\bfb$  (see also \cite{ibb}).  Although this term  can give something like $\p \bfd/\p t= \nabla\times \bfh$, it remains to properly define the meaning of $\bfd$ and $\bfh$.  Thus,   this bracket alone does not constitute  a closed theory.  Similarly  the coupling term (\ref{coupling}), a generalization of that introduced in \cite{morrison80,morrison82} that includes the internal variable $w$,  is written here in terms of $\bfd$, but the generalization of the Marsden-Weinstein term \cite{MW} (see also \cite{ibb2}) of (\ref{MW}),  and the first term of (\ref{pjm}), also a generalization of that given in  \cite{morrison80,morrison82},   are  unchanged.  The new internal term  here of (\ref{pjm}) with   $[\bar F_f,\bar{G}_f]_{w}$ does not depend on $\bfd$ and does not affect the Jacobi identity (cf.\ Appendix \ref{SpinJacobi}). 

To close the theory requires  a constitutive relation, something like  $\bfd=\tensepl\cdot\bfe$.  Such relations are often appended to electromagnetic theory based on phenomenological material properties, but here they emerge as a consequence of the Vlasov-like dynamics and the definitions (\ref{PM}).  Using (\ref{PM})  gives 
\bq
\bfd=\bfd[\bfe,\bfb;f]=\bfe + 4\pi \bfP[\bfe,\bfb;f]
\label{do}
\eq
with  both $\bfP$ and $\bfd$   linear in $f$, but not in $\bfe$ and $\bfb$.  In general these functionals  can be nonlinear and even global in nature.  It is only required that there be a unique inverse
\bq
\bfe=\bfd^{-1}[\bfd,\bfb;f]=\bfe[\bfd,\bfb;f]\,.
\label{eo}
\eq
Similarly,   using (\ref{PM}) 
\bq
\bfh=\bfh[\bfb,\bfe;f]=\bfb -4\pi \bfm[\bfb,\bfe;f]\,,
\label{ho}
\eq
which is also assumed to have an inverse, i.e.\ 
\bq
\bfb=\bfb[\bfh,\bfe;f] = \bfh + 4\pi\bfm[\bfh,\bfe;f]\,.
\label{bo}
\eq

For given $\calk$,  the expressions of (\ref{PM})  can be quite complicated, particularly when derivatives of the fields are included.  However,  these expressions are local in time, i.e.,  $\bfe$, $\bfb$, $\bfd$, and $f$ are all evaluated at the same time.  Because of the presence of $f$ and the equation governing it,  causal effects are included in a dynamical sense and do not need to be put in at the expense of breaking time-reversal symmetry.  Also,  there is no artificial separation of charge into bound and free components or current into magnetization or other.  Rather, charges and currents are determined dynamically according to the Vlasov equation.  A given charge may behave in any manner consistent with this dynamics.

When $\lim_{\bfe\rightarrow 0}\bfd[\bfe,\bfb;f]=0$ the first order term in an expansion in $\bfe$ gives $\bfd=\tensepl \cdot\bfe$, where $\tensepl$ is the dielectric permittivity operator.  Similarly,  $\bfb=\tensmu\cdot \bfh$, where $\tensmu$ is the permeability operator.  If one were to replace the Vlasov dynamics by trivial dynamics of linear response away from equilibrium, then one can recover the usual permittivity and permeability relations, including the usual causal (see e.g.\ \cite{ichi}) form in space and time.  But, this will not be done here.

To sum up,  the  Hamiltonian of (\ref{ham}) is given in terms of  $(\bfe,\bfb,f)$,  the  bracket of (\ref{pjm})-(\ref{dbkt}) in terms of $(\bfd,\bfb,f)$, and (\ref{do})  is a  closure relation relating $\bfd$ to $\bfe$.  Thus,  if  $\bar{H}[\bfd,\bfb,f]={H}[\bfe,\bfb,f]$ is defined by inserting the inverse of (\ref{do}) in for $\bfe$,  a closed theory is obtained.  For general $\calk$ this inversion cannot be done explicitly (although a series expansion may be possible); however,  the chain rule can be used to relate  functional derivatives of $H$ to those of $\bar{H}$ and thereby  obtain equations for the time derivatives of  $(\bfd,\bfb,f)$ which can then be shown to be  equivalent to those of Sec.~\ref{genkentheory}.   Alternatively,  the chain rule can be used to write the bracket of (\ref{pjm})-(\ref{dbkt}) in terms of the set of fundamental variables $(\bfe,\bfb,f)$.

To understand the chain rule,  suppose two functionals are related by
$
\bar F[\bfd,\bfb,f]= F[\bfe,\bfb,f]$,  where the right hand side is obtained by inserting (\ref{do}) in for $\bfd$ in the functional 
$F$ on the left. Variation of $\bar F=F$ gives
\bq
\int\!\!d\bfx\,\Big(\frac{\de \bar F}{\de \bfd}\cdot \de \bfd + \frac{\de \bar F}{\de \bfb}\cdot \de \bfb\Big) 
+ \int\!\! dz \, \frac{\de \bar F}{\de f}  \de f
= \int\!\!d\bfx\,\Big(\frac{\de   F}{\de \bfe}\cdot \de \bfe + \frac{\de F}{\de \bfb}\cdot \de \bfb\Big) 
+ \int\!\! dz \, \frac{\de  F}{\de f}  \de f\,,
\label{varF}
\eq
while variation of (\ref{eo}) gives 
\bq
\de\bfe=  
\frac{\de \bfe}{\de \bfd}\cdot \de \bfd+ \frac{\de \bfe}{\de \bfb}\cdot\de  \bfb +\frac{\de \bfe}{\de f} \de f\,, 
\label{epvar}
\eq
where ${\de \bfe}/{\de \bfd}$ etc.\ are  the usual Fr\'{e}chet derivatives obtained by first variation.    
Inserting  (\ref{epvar}) into (\ref{varF}) and comparing  the coefficients of the independent variations $\de \bfd$ etc.,   gives 
\bqy
\frac{\de \bar F}{\de \bfd}&=& \left(\frac{\de \bfe}{\de \bfd}\right)^{\dagger}\!\!\cdot \frac{\de   F}{\de \bfe}\,, 
\label{chainD}
\\
\frac{\de \bar F}{\de \bfb} &=& \frac{\de F}{\de \bfb} + \left(\frac{\de \bfe}{\de \bfb}\right)^{\dagger}\!\!\cdot \frac{\de   F}{\de \bfe}\,,
\label{chainB}
\\
\frac{\de \bar F}{\de f} &=&  \frac{\de   F}{\de f} + \left(\frac{\de \bfe}{\de f }\right)^{\dagger}\!\!\cdot \frac{\de   F}{\de \bfe}\,.
\label{chainf}
\eqy

A more explicit expression for $\left({\de \bfe}/{\de \bfd}\right)^{\dagger}$ can be obtained by varying (\ref{do}) at fixed $\bfb$ and $f$, giving
\bq
\de \bfd= \left(\tensi - 4\pi \frac{\de^2 K}{\de \bfe\de\bfe}\right) \cdot \de \bfe=:\tensep\cdot  \de \bfe
\label{eps}
\eq
or
\bq
\de \bfe= \frac{\de \bfe}{\de \bfd} \cdot \de \bfd=\tensep^{-1}\!\!\cdot \de \bfd
\eq
where $\tensep$ is the nonlinear permittivity operator (not to be confused with $\tensepl$).  Evidently,   
\bq
 \bfe_{\bfd} := \frac{\de \bfe}{\de \bfd}  = \tensep^{-1}  
 =\left(\tensi - 4\pi \frac{\de^2 K}{\de \bfe\de\bfe}\right)^{-1}\,,
 \label{e/d}
\eq
and $\bfe_{\bfd} = (\bfe_{\bfd})^{\dagger}$ or  $(\tensep^{-1})^{\dagger} =\tensep^{-1}$.  It is important to note 
that although $\de \bfd=\tensep\cdot \de \bfe$, $\bfd \neq \tensep \cdot \bfe$; the correct relation between $\bfd$ and $\bfe$ 
is given by the nonlinear expression of (\ref{do}).  Similarly,  
\bq
  \bfe_{\bfb} :=\frac{\de \bfe}{\de \bfb}= 4 \pi \,\tensep^{-1}\!\!\cdot  \frac{\de^2 K}{\de \bfb \de \bfe}  
   \quad {\rm and}\quad
 \bfe_{f} :=\frac{\de \bfe}{\de f}=4 \pi \, \tensep^{-1}\!\!\cdot  \frac{\de^2 K}{\de f \de \bfe} \,.
 \label{e/f}
 \eq
and the functional derivatives of (\ref{chainD})--(\ref{chainf}) can now be calculated:
 \bqy
 \bar{H}_{\bfd}&=& \bfe_{\bfd}^{\dagger} \cdot H_{\bfe}= \tensep^{-1}\!\cdot \tensep\cdot \frac{\bfe}{4\pi}
 \nonumber\\
 &=&\frac{\bfe}{4\pi}
  \nonumber\\\
  \bar{H}_{\bfb}&=& H_{\bfb} + \bfe_{\bfb}^{\dagger}\cdot H_{\bfe}= 
  K_{\bfb}  -\left(\frac{\de^2 K}{\de \bfb \de \bfe}\right)^{\dagger}\!\!\cdot \bfe +\frac{\bfb}{4\pi}
  + \left(\tensep^{-1}\!\!\cdot  \frac{\de^2 K}{\de \bfb \de \bfe} \right)^{\dagger} \!\!\cdot \tensep\cdot \bfe
  \nonumber\\
  &=& \frac{\bfb}{4\pi}-  \bfm
   \nonumber\\\ 
  \bar{H}_{f}&=&  H_f + \bfe_{f}^{\dagger}\cdot H_{\bfe}  
  =\calk - \left(\frac{\de^2 K}{\de f\de \bfe}\right)^{\dagger}\!\!\cdot \bfe 
  + \left(\tensep^{-1}\!\!\cdot  \frac{\de^2 K}{\de f \de \bfe} \right)^{\dagger}\!\!\cdot \tensep\cdot \bfe
  \nonumber\\
  &=& \calk\,.
  \label{DbarH}
  \eqy
Now,  using  the expressions of  (\ref{DbarH}) in  the bracket of  (\ref{pjm})--(\ref{dbkt}) gives
\bqy
\frac{\p \bfb}{\p t}&=& -4\pi c\, \nabla\times \bar{H}_{\bfd}= - c\,  \nabla\times \bfe
\label{dbdt}
\\
\frac{\p \bfd}{\p t}&=&4\pi c\,  \nabla\times \bar{H}_{\bfb} - \frac{4\pi e}{m} \int\!\!d\bfv  dw \, f\p_{\bfv} \bar{H}_f
 \nonumber\\
&=& c \,  \nabla\times \bfh  - \frac{4\pi e}{m} \int\!\!d\bfv  dw \, f\p_{\bfv}\calk   
\label{dddt}
\\
\frac{\p f}{\p t}&=& -\left[f,  \bar{H}_f \right]-  \p_{\bfv}\cdot\left(f \bar{H}_{\bfd}\right)
\nonumber\\
&=& -\left[f, \, \calk \right] - \frac{e}{m}\,  \bfe\cdot\frac{\p f}{\p \bfv} \,,
\eqy
where $[\,,\,]=[\,,\,]_{L} + [\,,\,]_w$ as defined by (\ref{nc1}) and (\ref{nc2}).  Thus the bracket reproduces  the Vlasov-like  equation of (\ref{genken}) and Maxwell's equations with the polarization and magnetization currents, but remember $f$ and $[\,,\,]$ can be written in terms of $\bfp$ using (\ref{v-p}) and thus the above is also equivalent to (\ref{genken}).

In the above, $\bfd$ is a convenience, a shorthand for  $\bfe + 4\pi \bfP$,  with $\bfe$ being the fundamental variable.  One could eliminate $\bfd$ from these equations e.g.\ by writing (\ref{dddt})   as follows:
\bq
\frac{\p \bfd}{\p t}= \bfe_{\bfd} \cdot \frac{\p \bfe}{\p t} + \bfP_{\bfb}\cdot \frac{\p \bfb}{\p t} + 
\bfP_{f}\cdot \frac{\p  f}{\p t} \,,
\label{complicated2}
\eq
where $\bfe_{\bfd}=\tensep$, as before,  and  $\bfP_{\bfb}$  and $\bfP_{f}$  are  again operators obtained by variation of $\bfP$,  and then inserting the other two equations of motion for the time derivatives.  This procedure will lead to a complicated set of equations in terms of the fundamental variables $(\bfe,\bfb,f)$.  
Another way of obtaining these complicated equations is to obtain a bracket in terms of $\bfe,\bfb$, and $f$ alone, by inserting the transformations for the functional derivatives of (\ref{chainD}), (\ref{chainB}), and (\ref{chainf}) into the bracket of (\ref{pjm})--(\ref{dbkt}).  This yields the following complicated bracket:
 \bqy
\{F,G\}&=&\int\!\!d\bfx d\bfv  dw\, 
f  
\left[{F}_f + \bfe_{f}^{\dagger}\!\cdot {F}_{\bfe}\, , \,  {G}_f + \bfe_{f}^{\dagger}\! \cdot {G}_{\bfe}\right] 
\nonumber\\
&+&   \frac{4\pi e}{m}\int\!\!d\bfx d\bfv  dw\, f\left( \left(\bfe_{\bfd}^{\dagger}\cdot  {G}_{\bfe}  \right)
         \cdot   \p_{\bfv}
    \left( {F}_f +  \bfe_{f}^{\dagger}\cdot {F}_{\bfe}\right) 
-  \left(\bfe_{\bfd}^{\dagger}\cdot  {F}_{\bfe} \right)
           \cdot\p_{\bfv} 
     \left( {G}_f + \bfe_{f}^{\dagger}\cdot {G}_{\bfe}\right)
     \right)
\nonumber\\
&{\  }& \hspace{.5 in} +4\pi c  \int\!\!d \bfx \,
\Big(
\left(\bfe_{\bfd}^{\dagger}\cdot  {F}_{\bfe}\right)
\cdot \nabla\times  \left(  {G}_{\bfb} +   \bfe_{\bfb }^{\dagger}\!\!\cdot  {G}_{\bfe}\right) 
\nonumber\\
&{\  }& \hspace{1.5 in}  
-  \left(\bfe_{\bfd}^{\dagger}\cdot  {G}_{\bfe}\right)
\cdot \nabla\times  \left(  {F}_{\bfb} +   \bfe_{\bfb }^{\dagger}\!\!\cdot  {F}_{\bfe}\right) 
\Big)\,,
\label{ebkt}
\eqy
where to complete the procedure the expressions of (\ref{e/d}) and (\ref{e/f}) are to be  inserted. 
 Thus the  bracket of (\ref{ebkt}) is quite complicated with many terms and operators.  Nevertheless, by its construction  {\it it satisfies the Jacobi identity} (modulo the $\nabla\cdot\bfb=0$ obstruction discussed in Appendix \ref{MVjacobi}).   With Hamiltonian $H[\bfe,\bfb,f]$  it is easily seen from (\ref{DbarH}) that this bracket  produces the correct equations for $\p f/\p t$ and $\p \bfb/\p t$,  but it is less easy, yet possible,  to see it produces the complications of (\ref{complicated2}) correctly.

\medskip

For some theories of interest, including usual linear response theory,  $\calk$ has a  simplified form, viz.
\bq
\calk(\bfv,w, \bfe,\bfb)= h(\bfv,w,\bfb) +\mathbf{\calp}(\bfv,w,\bfb)\cdot \bfe 
+ {\frac1{2}}\, \bfe\cdot\tensk(\bfv,w,\bfb)\cdot \bfe\,,
\label{quadk}
\eq
where $\tensk^{\dagger}=\tensk$.  For example, such a  linear polarization theory is sufficient for some drift and gyrokinetic-like theories. 
With (\ref{quadk}) 
\bq
\bfP=-\int\!\!d\bfv dw\, \frac{\p \calk}{\p \bfe} f=- \int\!\!d\bfv dw\, \mathbf{\calp} f 
-\bfe\cdot \int\!\!d\bfv dw\, \tensk\, f\,.
\label{p2}
\eq
The first term of the second equality of (\ref{p2}) represents a permanent dipole moment per unit volume, which   can be dropped: because  (\ref{ham}) is a Legendre transform it  will cancel out anyway.  The second term of (\ref{p2}) defines the electric susceptibility, $\tensche$.  Thus $\bfP= \tensche\cdot \bfe$ or
\bq
\bfd=\tensepl \cdot \bfe
\eq
with $\tensepl =  \tensi + 4\pi \tensche$.  Note that unlike the $\tensep$ of (\ref{eps}), when  (\ref{quadk}) is assumed,  
$\tensepl$ is independent of $\bfe$ but not of $\bfb$.   Explicitly  in terms of indices
\bq
\epsilon_{ij}=\de_{ij} -4 \pi  \int \!\!  d\bfv dw\, k_{ij}(\bfv,w,\bfb)\, f   \,.
\label{epten}
\eq
Assuming   $\tensepl^{-1}$ exists and has the form
\bq
\tensepl^{-1}= \frac{\tensi}{\tensi+4\pi \tensche}= \tensi- 4\pi \tensche  +(4\pi \tensche)^2 - \dots\,,
\eq
where $\tensche^2$ stands for matrix multiplication, and so on down the line.  Hence,   $\bfe=\tensepl^{-1}\cdot \bfd$. 
Although $\tensepl$ is linear in $f$, this is not the case for $\tensepl^{-1}$.  Finally,  for the $\calk$ of (\ref{quadk}) that is quadratic in $\bfe$,   the following simplified expressions are obtained:
 \bqy
 \bfe_{\bfd}^{\dagger}&=& \tensepl^{-1}
 \\
  \bfe_{\bfb}^{\dagger}&=& \bfd\cdot\frac{\p\,  \tensepl^{-1}}{\p\bfb}
 \\
  \bfe_{f}^{\dagger}&=&  \bfd \cdot \tensepl^{-1}\!\cdot \tensk\cdot  \tensepl^{-1} \,.
 \eqy 
 Note, obtaining these formulas can be facilitated by using  identities obtained by varying the expression $\tensepl^{-1}\cdot\tensepl =\tensi$.

 
 \section{Examples}
 \label{examples}
 
 In this section four examples are given: that of Sec.~\ref{VM} is the usual Vlasov-Maxwell theory, that of Sec.~\ref{GCDKT} is a guiding center drift kinetic theory that includes nontrivial polarization and magnetization effects, that of Sec.~\ref{spin} includes a physically perspicuous internal variable, and that of Sec.~\ref{monopole} was chosen to show the generality of the lift theory by altering Maxwell's equations.

\subsection{Vlasov-Maxwell}
\label{VM}

For Vlasov-Maxwell theory $w$ is nonexistent and only $z=(\mathbf{x}, \mathbf{p})$ appears. 
Thus, $\bar f(\mathbf{x}, \mathbf{p},t)$ and, with  $\calk={|\mathbf{p}-e\mathbf{A}/c|^2}/{2m}$, 
Eq.~(\ref{genken}) becomes
\bqy
 \frac{\p \bar f}{\p t}&=&\left[\frac{|\mathbf{p}-e\mathbf{A}/c|^2}{2m} + e\phi,\bar f\right]
 \nonumber\\
 &=&- \frac{e}{mc}\left(\mathbf{p}-e\mathbf{A}/c\right)
\cdot  \nabla \mathbf{A} \cdot    \frac{\p \bar f}{\p \mathbf{p}}
+ e\nabla \phi \cdot\frac{\p \bar f}{\p \mathbf{p}}
-\frac1{m} \left(\mathbf{p}-e\mathbf{A}/c\right)\cdot \nabla \bar f\,.
\label{fA}
 \eqy
 In terms of $f(\bfx,\bfv,t)$,   $\calk=m|\bfv|^2/2$ and (\ref{fA}) becomes
 \bq
   \frac{\p f}{\p t}=  -   \mathbf{v}\cdot \nabla  {f}- \frac{e}{m}\,\left(\mathbf{E} +\frac{\mathbf{v}}{c}\times \mathbf{B}\right)\cdot \frac{\p   f}{\p  \mathbf{v}}\,.
   \label{usualVM}
 \eq

From the general Hamiltonian (\ref{ham}),  with  $\calk=m|\bfv|^2/2$,   evidently the Vlasov-Maxwell Hamiltonian is  
\bq
H=\frac{m}{2}\int \! d\bfx d\bfv \, |\bfv|^2 \, f + \frac1{8\pi } \int \!d\bfx \left(|\bfe|^2 + |\bfb|^2\right)\,, 
\label{VMham}
\eq
and with $[g,h]_L$ the bracket of (\ref{pjm})--(\ref{dbkt})  becomes  the  Vlasov-Maxwell bracket
\bqy
\Brac FG&=&\int \! d\bfx d\bfv \,\left( f\brac{F_f}{G_f}_{\bfv} + f\brac{F_f}{G_f}_{\bfb} 
+\frac{4\pi e}{m}\, f \left(G_{\bfe}\cdot\p_{\bfv}F_f - F_\bfe \cdot\p_{\bfv} G_f\right)\right) \nonumber\\
&{\ }& \hspace{2 cm} +4\pi c  \int\!d\bfx \, \left(F_{\bfe}\cdot \nabla\times G_{\bfb} - G_{\bfe}\cdot\nabla\times F_{\bfb}\right)\,. 
 \label{MVbkt}
\eqy
With this bracket and Hamiltonian,  one obtains  the usual  Vlasov-Maxwell equation as 
 \bq
  \frac{\p f}{\p t}= \{f,H\}= - \left[f,\calk\right]_{L}- \frac{e}{m}\, \bfe\cdot \frac{\p   f}{\p  \mathbf{v}}\,,
\eq
which is equivalent to (\ref{usualVM}).  Similarly, since for  this example $\bfd=\bfe$,   the usual expression for the current is obtained from $\{\bfe,H\}$, there being no polarization or magnetization contributions, and Faraday's law is given by  $\p \bfb/\p t=\{\bfb,H\}$.

The relativistic Vlasov-Maxwell theory similarly follows with the choice
$
 \calk=c\sqrt{|\bfp-e\bfa/c|^2 + m_0^2c^2}
$
and the theory can be written in manifestly covariant form \cite{PM85}, but  this will not pursued  further here. 
 
\subsection{Guiding Center Drift Kinetic Theory}
\label{GCDKT}

A canonical Hamiltonian description for guiding center particle motion was obtained in \cite{PM85,PM91} by applying Dirac constraint theory to Littlejohn's degenerate Lagrangian \cite{littlejohn81,littlejohn83} (with a regularization suggested in \cite{wimmel}) in order to effect a Legendre transformation.  The canonical variables of the theory are $(\bfx,q_4,\bfp,p_4)$ and the particle Hamiltonian,  Dirac's primary Hamiltonian for this problem (see e.g.\ \cite{sudarshan,PM91}),  is
\bqy
  \cale(\bfx,q_4,\bfp,p_4,\bfe,\bfb,\nabla \bfb)&=&\bfv_g\cdot(\bfp-e\bfa^*/c) + V_4p_4 + e\phi^*
\\
&=&   \calk\big(\mathbf{p}- {e\mathbf{A}}/{c},q_4,p_4, \mathbf{E},\mathbf{B}, \nabla\bfb\big)+ e\phi\,,  
\label{gccale}
\eqy
which is of the form of (\ref{gHam}).     Here
\bqy
\bfa^*&=& \bfa + \ub \, mcv_0\ga(q_4/v_0)/e\,, 
\qquad
e\ph^*=e\ph+  \mu |\bfb| + m\left(q_4^2 + |\bfv_{E}|^2\right)/2\,, 
\\
\bfv_{E}&=&c\left(\bfe\times\bfb\right)/|\bfb|^2\,,
\qquad
\ub=\bfb/|\bfb| \,,
\\
\bfv_g&=& q_4\bfb^*/(\ga'B_{||}^*) +c\bfe^*\times \bfb/B_{||}^*\,,
\qquad
V_4= e\bfe^*\cdot\bfb^*/(mg'B_{||}^*)\,, 
\\
\bfb^*&=&\nabla\times\bfa^*
\qquad {\rm and}\qquad 
\bfe^*=-\nabla\phi^*-\frac1{c}\frac{\p \bfa^*}{\p t}
\,.
\eqy
The function $\ga(z)$ is an antisymmetric regularization function with $z=q_4/v_0$ and $v_0$ 
some constant velocity.  The Littlejohn theory is recovered if $\ga(z)=z$, in which case 
$q_4=v_{||}$.  In the regularized theory $\ga(z)\approx z$ for small $|z|$, but 
for large $|z|$, $\ga$ is bounded so that with $v_0>>v_{thermal}$,
$mv_0c\,  \ga(\infty)<<e|\bfb|/({\ub\cdot\nabla\times\ub})$, 
which is accomplished, e.g.,  by $\ga=\tanh(z)$.

This theory has an eight-dimensional phase space with the canonical bracket
\bq
[g,h]=[g,h]_{\bfp} + {\p_{q_4}}g\, {\p_{p_4}}h -   {\p_{q_4}}h\, {\p_{p_4}}g\,, 
\label{8bkt}
\eq
which with  $\calk$ of (\ref{gccale}) completes the theory.  The appropriate bracket of the form of  (\ref{pjm})--(\ref{dbkt}) is obtained  with (\ref{8bkt}), and from  $\calk$ the functional $K$ can be constructed and thus the Hamiltonian of (\ref{ham}). This bracket and Hamiltonian produces the equations of motion;  thus this system is a Hamiltonian field theory.  From  $K$  the polarization and magnetization are can be obtained straightforwardly.   Since expressions are complicated, the reader is referred to  \cite{PM85,PM91} for details.
  
\subsection{Spin Vlasov-Maxwell}
\label{spin}

The nonrelativistic  spin Vlasov-Maxwell system is a kinetic theory generalization of the Vlasov-Maxwell system that includes  a  semiclassical description of spin    \cite{brodmark,brodmark2,markmor}.  The Hamiltonian description of this system \cite{markmor} will be shown to fit  within the gauge-free lifiting framework.   The  spin Vlasov-Maxwell electron  distribution function,   $f(\bfx,\bfv,\bfs,t)$, contains the internal spin variable $\bfs=(s_1,s_2,s_3)$, and satisfies 
\bq
\frac{\p f}{\p t} = -\bfv \cdot\nabla f 
+ \left[ \frac{e}{m}\left(\bfe +\frac{\bfv}{c}\times\bfb\right) +\frac{2\mu_e}{m \hbar c} \nabla(\bfs\cdot\bfb)\right]\cdot\frac{\p f}{\p \bfv} +\frac{2\mu_e}{\hbar c}\left(\bfs\times\bfb\right)\cdot \frac{\p f}{\p \bfs}\,,
\label{smv}
\eq
where $m$ and $e>0$ are  the electron mass and charge, respectively, $2\pi\hbar$ is Planck's constant,   $\mu_e=g\mu_{\bfb}/2$  is the electron magnetic moment in terms of   $\mu_{\bfb}$,  the Bohr magneton,   and the electron spin $g$-factor.   Equation (\ref{smv}) is coupled to the dynamical Maxwell's equations, 
\bq
\frac{\p \bfb}{\p t}=-c\, \nabla\times \bfe\,,
\qquad
\frac{\p \bfe}{\p t}=c\, \nabla\times \bfb-4\pi \bfj\,,
\label{faramp}
\eq
 through  the current $\bfj=\bfj_f+c\nabla \times \bfm$, which  has ``free'' and spin magnetization parts:
\bq
\bfj_f:=-e\int\!\!d\bfv\,d\bfs \,\, \bfv  f
\qquad {\rm and}\qquad
\bfm:=-\frac{2\mu_e}{\hbar} \int\!\!d\bfv\,d\bfs \,\,\bfs f \,.
\label{mag}
\eq
Equations~(\ref{smv})  and (\ref{faramp}), with  (\ref{mag}),  are to be viewed classically and consequently a full nine-dimensional phase space integration,  $d^9z=d\bfx d\bfv d\bfs$,  is considered for $f$.  Spin quantization is obtained as an initial condition that constrains $\bfs$ to lie on a sphere (see \cite{markmor}).  Extension to multiple species is straightforward, so will not be included. 

For this system $\calk$ is chosen as follows:
\bq
\calk(\bfv,\bfs, \bfb)=\frac{m}{2} |\bfv|^2 + \frac{2\mu_e}{\hbar c} \bfs \cdot \bfb\,,
\label{spincalk}
\eq
however,  more general forms are possible.  It is not difficult to check that the characteristic equations of (\ref{smv}) are of the form of  (\ref{genken}) with the Poisson bracket $[g,h]= [g,h]_{\bfv} + [g,h]_{\bfs}$, 
where  
\bq
\brac gh_{\bfs}=    \bfs\cdot (\p_{\bfs} g\times \p_{\bfs} h)\,, 
\label{pspin}
\eq
with $\p_{\bfs}:={\p}/{\p \bfs}$, and $\calk$ given by (\ref{spincalk}).

Thus  the  Hamiltonian functional (\ref{ham}) becomes  
\bq
H[\bfe,\bfb,f]=\int\!\!d^9z\, \left(\frac{m}{2} |\bfv|^2 + \frac{2\mu_e}{\hbar c} \bfs \cdot \bfb\right) f + \frac1{8\pi}\int\!\! d\bfx\, \left(|\bfe|^2 + |\bfb|^2\right)\,,
\eq
which can be shown to be conserved directly by using the equations of motion,  and the bracket of (\ref{pjm})--(\ref{dbkt}) adapted to the present example is   
\bqy
\Brac FG_{sVM}&=&\int \! d^9z  \, f \Big(\brac{F_f}{G_f}_{\bfv}  +  \brac{F_f}{G_f}_{\bfb}  +  \brac{F_f}{G_f}_{\bfs} 
\label{new}\\
 &{\ }& \hspace{ .7 in}
 + \frac{4\pi e}{m}\left(F_{\bfe}\cdot\p_{\bfv} G_f - G_{\bfe}\cdot\p_{\bfv} F_f\right) \Big)
  \label{couple}\\
&{\ }& \hspace{ 1.2 in} +4\pi c  \!\int\!d\bfx \, \left(F_{\bfe}\cdot \nabla\times G_{\bfb} - G_{\bfe}\cdot\nabla\times F_{\bfb}\right)\,. 
 \label{MVSbkt}
\eqy

 The last term  of (\ref{new}) of $\{\,,\,\}_{sVM}$  accommodates the spin, an internal variable; it is not surprising that it has an inner bracket based on the $\mathfrak{so}(3)$ algebra (e.g.~\cite{sudarshan}).  The remaining terms  of (\ref{new}), (\ref{couple}),  and (\ref{MVSbkt})   produce the usual terms of Vlasov-Maxwell theory: it  is a straightforward  exercise to show that Eqs.~(\ref{smv}) and (\ref{faramp}) are given as follows:
 \bq
 \frac{\p f}{\p t}=\{f,H\}_{sVM}
 \,, \qquad
 \frac{\p \bfb}{\p t}=\{\bfb,H\}_{sVM}
  \,, \qquad
 \frac{\p \bfe}{\p t}=\{\bfe,H\}_{sVM}\,.
  \nonumber
  \eq
This is facilitated by the identity $\int d^9z\,  f[g,h]=-\int d^9z\, g[f,h]$, which works for each of $\brac gh_{\bfv}$, $\brac gh_{\bfb}$, and $\brac gh_{\bfs}$.  It follows that the polarization $\bfP\equiv 0$ and $\de K/\de \bfb=-\bfm$ as given by (\ref{mag}).

In  Appendix \ref{SpinJacobi} a direct proof of  the Jacobi identity for $\{f,H\}_{sVM}
$ is given.

\subsection{Monopole Vlasov-Maxwell}
\label{monopole}

This example differs from the previous ones in that Maxwell's equations are changed to another field theory.  In particular, Dirac's theory of electromagnetism \cite{dirac1,dirac2} with monopole charge will be treated in the lift framework.

For Dirac's  theory,  (\ref{spliteom}) is replaced by the particle orbit equations
\bq
\dot{\mathbf{x}}= \bfv \qquad {\rm and}\qquad \dot\bfv=\frac{e}{m}( \bfe +\frac{\bfv}{c}\times \bfb)
+ \frac{g}{m}( \bfb -\frac{\bfv}{c}\times \bfe)\,.
\eq
where $g$ and $e$ are magnetic and electric charges, respectively.   The appropriate particle Poisson bracket for insertion into (\ref{pjm})--(\ref{dbkt}) is
\bq
[g,h]_m=[g,h]_{\bfv}+ [g,h]_{\bfb} + [g,h]_{\bfe}
\label{monobkt}
\eq
where
\bq 
\brac fg_{\bfe}:=-\frac{g}{m^2c}\bfe\cdot\left(\p_{\bfv} f\times\p_{\bfv}  g\right)
\label{Ebkt}\,.
\eq
The particle Hamiltonian is given by $\calk=m|\bfv|^2/2$, just as for Vlasov-Maxwell theory.

In lifting to a kinetic theory there are various kinds of multi-species dynamics,  with and without magnetic charge, that could be considered.   Here the case of  a  single species of identical particles that carry both magnetic and electric charges will be developed, along the lines of the quantum fluid theory considered in \cite{ibbdirac}.  
Inserting (\ref{monobkt}) into (\ref{pjm})--(\ref{dbkt}) and adding a new coupling term to account for the $\bfb$ that acts as an external force, gives  the bracket
\bqy
\Brac FG_{mVM}&=&\int \! d\bfx d \bfv  \, \Big(
f\brac{F_f}{G_f}_m 
 \\
&{\ }& \hspace{.5 cm} +\frac{4\pi e}{m}\, f \left(G_\bfe\cdot\p_{\bfv} F_f - F_{\bfe}\cdot\p_{\bfv} G_f\right)  
+  \frac{4\pi g}{m}\, f \left(G_{\bfb}\cdot\p_{\bfv}F_f - F_{\bfb}\cdot\p_{\bfv} G_f\right) \Big)
\\
&{\ }& \hspace{2 cm} + 4\pi c \int\!d\bfx \, \big(F_{\bfe}\cdot \nabla\times G_{\bfb} - G_{\bfe}\cdot\nabla\times F_{\bfb}\big)
\,.
\label{MMVbkt}
\eqy
With  Hamiltonian of  (\ref{VMham}) this bracket yields
\bqy
\frac{\p f}{\p t}&=& - \bfv\cdot\nabla f -\frac{\p f}{\p \bfv}\cdot
\left(
\frac{e}{m}( \bfe + \frac{\bfv}{c}\times \bfb)
+ \frac{g}{m}( \bfb - \frac{\bfv}{c}\times \bfe)
\right)
\label{ft}\\
\frac{\p \bfe}{\p t}&=&c\, \nabla\times \bfb -4\pi \bfj_e
\label{et}\\
\frac{\p \bfb}{\p t}&=& -c\, \nabla\times \bfe -4\pi \bfj_m
\label{bt}
\eqy
where
\bq
\bfj_e=e\int f v \, d\bfx \qquad {\rm and }\qquad
\bfj_m=g\int f v \, d\bfx \,.
\eq
Thus monopole Vlasov-Maxwell is a Hamiltonian field theory. 

One reason for investigating Dirac's model in the present context is to  see if the $\nabla\cdot\bfb=0$ obstruction to Jacobi discussed in Appendix \ref{Ajacobi} can be removed.  In Appendix \ref{MonopoleJacobi} the Jacobi identity for 
$\{F,G\}_{mVM}$ is proved directly, and there it is discovered that the solenoidal character of $\bfb$ is replaced by
$\nabla\cdot (e\bfb- g\bfe )=0$.  Thus, the space of functionals must still be restricted to such fields, as discussed in Appendix \ref{Ajacobi}.  However,   Dirac constraint theory can reduce this to a boundary condition at infinity \cite{taint}. 

Another reason for investigating monopole theories is for their utility in developing numerical algorithms. For example, the Gudunov numerical method for magnetohydrodynamics \cite{gudunov} (see also \cite{dellar}) exploits  a form that allows for $\nabla\cdot\bfb\neq0$ that was subsequently shown to be Hamiltonian without the $\nabla\cdot\bfb=0$ constraint in \cite{morgreene,morrison82}  (see also \cite{amp0,amp1,amp2}).  However, since the monopole theory of this section  requires  a specific linear combination of $\bfe$ and $\bfb$ to be divergence free, adaptations of these methods in not  so straightforward.    In any event, the reader can rest assured  if mononpoles are discovered there still will exist Hamiltonian guiding center and gyrokinetic kinetic theories, obtained with suitable choices for $\calk$ with associated generalized polarization and magnetization vectors.


\section{Conclusions}
\label{conclusions}

The main accomplishment of this work is to describe how a matter model of dynamics can be lifted to a Hamiltonian coupled Vlasov-Maxwell  system.  En route to this Hamiltonian theory,  the general constitutive relations of (\ref{eo}) and (\ref{ho}) or, equivalently,  the nonlinear permittivity and permeability operators,   as determined by (\ref{PM}), were obtained.  These  constitutive functionals  are very general: as discussed in the paper,   $\calk$ may contain all derivatives of the fields and  may  even be global in nature and contain integral operators.  From the general constitutive functionals,  it was shown how to obtain the usual linear relations.  A noncanonical Poisson bracket, another step in the program started in  \cite{morgreene,morrison80,morrison82}, was obtained for this general class of theories.   

Four examples were given, including the general class of guiding center kinetic theories of Sec.~\ref{GCDKT}.   This latter example, like all systems that are in the class of variational theories of \cite{PM85}, easily was shown to possess  the Hamiltonian structure.  Thus, the theory of this paper determines  the  path to follow for  obtaining the Hamiltonian formulation of a consistent gyrokinetic theory by making use of the results  \cite{pfrischGK1,pfrischGK2}.  
 
Various generalizations are possible.  Namely, the extension to many species of different dynamics,  relativistic theory other than the Vlasov-Maxwell example that was described,  versions where  particle matter models have more general finite-dimensional Poisson brackets, are all straightforward.  Also,  extending the matter dynamics by coupling to other general  gauge-free field theories is possible.  

Another application of the  techniques of this paper will be used in a subsequent work  \cite{morrison12} that will treat Hamiltonian perturbation theory in the field theory context.  There it will be shown how an exact transformation of the particle (characteristic) equations of Vlasov-Maxwell equations can be lifted to the kinetic and Maxwell equations, and how this can be used in perturbation theory for infinite-dimensional noncanonical Hamiltonian systems. 

In physics there are two  ways of constructing new theories.  The usual way is to construct an action principle by postulating a Lagrangian density with the desired observables and symmetry group properties.   Alternatively one can postulate an energy functional and Poisson bracket, which is essentially the approach of the present paper.   With this latter approach, one must prove directly the Jacobi identity $\nest FGH + \nest GHF+ \nest HFG \equiv 0$,  for all functionals $F$, $G$, and $H$.  Techniques for doing this are not generally known, and this provides one reason for  Appendix \ref{Ajacobi}.

\appendix


\section{Direct proof of Jacobi identities}
\label{Ajacobi}

One  term of the original Vlasov-Maxwell Hamiltonian formulation of \cite{morrison80} presented an obstruction to the Jacobi identity \cite{weinmor}.  This term was replaced  in \cite{MW} in order remove this problem, but it was then reported in \cite{morrison82} that the new term also presents an obstruction, viz.\ that  given by (\ref{main}) below.
One can rescue the Hamiltonian theory by requiring all functionals to depend on fields  $\bfb$ such that  $\nabla\cdot\bfb=0$, but because the orginal program begun in  \cite{morgreene,morrison80} was to  construct  truly gauge-free field theories in terms of noncanonical Poisson brackets, this taint was a disappointment.  The same obstruction appeared in the context of magnetohydrodynamics \cite{morrison80}, but a way to remove it was obtained in \cite{morrison82}.  To date, the best fix for the taint of the Vlasov-Maxwell bracket is given in  \cite{taint} by using Dirac constraint theory, which replaces $\nabla\cdot\bfb=0$ by a boundary condition at infinity.

Appendix \ref{MVjacobi} contains the details of the onerous calculation first performed by the author  in 1981 (reported in detail here for the first time, as it was originally done),  leading to the result of (\ref{main}) that  appeared in \cite{morrison82}.   This is followed in Appendices \ref{SpinJacobi} and \ref{MonopoleJacobi} by a direct proof of the brackets for the spin and monopole Vlasov-Maxwell theories, respectively.

\subsection{Jacobi identity for the Vlasov-Maxwell bracket}
\label{MVjacobi}

For convenience the charge, mass, and a factor of $4\pi$ are scaled out to obtain the Vlasov-Maxwell bracket for the fields $f(z,t)$, $E(\bfx,t)$, and $B(\bfx,t)$ in the following form:
\bqy
\Brac FG&=&\int \! d^6z \, f\brac{F_f}{G_f}_c + f\brac{F_f}{G_f}_B 
+f \left(G_E\cdot\p_vF_f - F_E\cdot\p_v G_f\right) \nonumber\\
&{\ }& \hspace{2 cm} +\int\!d^3x \, F_E\cdot \nabla\times G_B - G_E\cdot\nabla\times F_B
 \nonumber\\
 &=:&\Brac{F}{G}_c + \Brac{F}{G}_B +\Brac{F}{G}_{Ef}+\Brac{F}{G}_{EB}\,,
 \label{MVbktS}
\eqy
where $F_f:= {\de F}{/\de f}$, $\brac fg_c:=\nabla f\cdot\p_vg - \nabla g\cdot\p_vf$, $\p_v:= {\p}/{\p v}$,  
$\brac fg_B:=B\cdot\left(\p_v f\times\p_v g\right)$, 
and $\Brac{F}{G}_c$ etc. are obvious from context.  Also, note, boldface has  been removed  since the formulas are busy enough.   Since charge can be scaled out in this manner, it is evident that the validity of the Jacobi identity is independent of the sign of the species charge.  
 
 Above, the term  $\Brac{F}{G}_B$,  the Marsden-Weinstein term \cite{MW}, has been  separated out  because it will be seen to be the source of the failure of Jacobi identity unless $\nabla \cdot \bfb=0$.  Considering the combination $\Brac{F}{G}_c + \Brac{F}{G}_B$ together  would simplify the calculation somewhat.

The Jacobi identity is
\bqy
 \nest FGH&:=:&\Brac{\Brac{F}{G}_c}H + \Brac{\Brac{F}{G}_B}H +
 \Brac{\Brac{F}{G}_{Ef}}H + \Brac{\Brac{F}{G}_{EB}}H\nonumber\\
  &=&
  \underset{1}{\underbrace{\Brac{\Brac{F}{G}_c}H_c}}
 + \underset{2}{\underbrace{\Brac{\Brac{F}{G}_c}H_B}}
 + \underset{3}{\underbrace{\Brac{\Brac{F}{G}_c}H_{Ef}}}
  +\underset{4}{\underbrace{\Brac{\Brac{F}{G}_c}H_{EB}}}
   \nonumber\\
    &+&
  \underset{5}{\underbrace{\Brac{\Brac{F}{G}_B}H_c}}
 + \underset{6}{\underbrace{\Brac{\Brac{F}{G}_B}H_B}}
 + \underset{7}{\underbrace{\Brac{\Brac{F}{G}_B}H_{Ef}}}
  +\underset{8}{\underbrace{\Brac{\Brac{F}{G}_B}H_{EB}}}
   \nonumber\\
    &+&
   \underset{9}{\underbrace{\Brac{\Brac{F}{G}_{Ef}}H_c}}
 + \underset{10}{\underbrace{\Brac{\Brac{F}{G}_{Ef}}H_B}}
 + \underset{11}{\underbrace{\Brac{\Brac{F}{G}_{Ef}}H_{Ef}}}
  +\underset{12}{\underbrace{\Brac{\Brac{F}{G}_{Ef}}H_{EB}}} \nonumber\\
      &+&
   \underset{13}{\underbrace{\Brac{\Brac{F}{G}_{EB}}H_c}}
 + \underset{14}{\underbrace{\Brac{\Brac{F}{G}_{EB}}H_B}}
 + \underset{15}{\underbrace{\Brac{\Brac{F}{G}_{EB}}H_{Ef}}}
  +\underset{16}{\underbrace{\Brac{\Brac{F}{G}_{EB}}H_{EB}}} \nonumber\,,
  \eqy
where the symbol $:=:$ means modulo cyclic permutation;    `cyc'  will be  used to denote cyclic permutation when only one side is to be permuted.

To prove the Jacobi identity for a generic Poisson bracket of the form 
$\Brac{F}{G}=\langle F_{\psi}|\calj G_{\psi}\rangle$, with cosymplectic operator $\calj$, one must calculate 
the functional derivative $\de\Brac{F}{G}/\de \psi$.  This derivative has three contributions: 
one from $F_\psi$, one from $\calj$, and one from $G_{\psi}$.   The first and last give rise to second functional derivatives.  Proofs of the Jacobi identity are greatly simplified by  the following:

\medskip

\noindent{\bf Bracket Theorem}  (\cite{morrison82}) {\it To prove the Jacobi identity for generic brackets of the form $\Brac{F}{G}=\langle F_{\psi}|\calj G_{\psi}\rangle$ one need only consider the explicit dependence of $\calj$ on $\psi$ when taking the functional derivative $\de\Brac{F}{G}/\de \psi$. }

\proof The formal proof uses the anti-self-adjointness  of $\calj$ and the self-adjointness of the second functional derivative.  With these symmetries it can be shown that all second functional derivative terms cancel.\ $\Box$

\medskip

In what follows   $\de\Brac{F}{G}/\de \psi \deq \, \dots$  denotes the functional derivative modulo the second derivative terms.  Equation (\ref{MVbktS}) gives 
\bqy
\frac{\de \Brac{F}{G}_{c}}{\de f}&\deq&  \brac{F_f}{G_f}_c \,,
 \quad \frac{\de \Brac{F}{G}_{c}}{\de B}\deq 0\,,
 \quad \frac{\de \Brac{F}{G}_{c}}{\de E}\deq 0\,,
 \label{derC}\\
\frac{\de \Brac{F}{G}_{B}}{\de f}&\deq&   \brac{F_f}{G_f}_B\,, 
\quad \frac{\de \Brac{F}{G}_{B}}{\de B}\deq \! \!\int\!\! dv\, f  (\p_vF_f\times\p_v G_f)
\,,\quad \frac{\de \Brac{F}{G}_{B}}{\de E}\deq 0\,,
 \label{derB}\\
\frac{\de \Brac{F}{G}_{Ef}}{\de f}&\deq& \left(G_E\cdot \p_v F_f - F_E\cdot \p_v G_f\right)\,,  
\quad \frac{\de \Brac{F}{G}_{Ef}}{\de B}\deq 0\,,
\quad \frac{\de \Brac{F}{G}_{Ef}}{\de E}\deq 0\,,
 \label{derEF}\\
\frac{\de \Brac{F}{G}_{EB}}{\de f}&\deq& 0\,,
\quad \frac{\de \Brac{F}{G}_{EB}}{\de B}\deq 0\,,
\quad \frac{\de \Brac{F}{G}_{EB}}{\de E}\deq 0\,.
 \label{derEB}
\eqy
  
 The following are immediate:
 
 \begin{itemize}
 \item \underline{Term 1} vanishes because $\Brac{F}{G}_c$  is Lie-Poisson; i.e.,  using the first of (\ref{derC}), 
 $\de \Brac{F}{G}_c/\de f\deq\brac{F_f}{G_f}$,  which,  when inserted into $\Brac{\Brac{F}{G}_c}H_c$ and cyclicly permuting,  vanishes   by virtue of the Jacobi identity of $[\ ,\ ]_c$. 
 \item \underline{Term 4}  vanishes by the Bracket Theorem because of the second two equations of (\ref{derC}).
 \item \underline{Term 12}  vanishes by the Bracket Theorem because of the second two equations of (\ref{derEF})
 \item\underline{Terms 13-16} vanish by the Bracket Theorem because   $\Brac{F}{G}_{EB}$ has  no explicit  dependence on $f$, $E$, or $B$, i.e.\ because of (\ref{derEB}). 
\end{itemize}

\noindent{\bf Remark.}  One can organize Jacobi identity calculations at  the outset by grouping together all like terms that can possibly cancel.  For example terms with the same functional derivatives of $F$, $G$, and $H$ must be considered together.  Sometimes other considerations can aid in the grouping of terms.  When terms are grouped appropriately,  failure of a class of terms to cancel is a proof of the failure of Jacobi.  In the heading below $fff$ means that only function derivatives with respect to $f$ occur, etc.

\medskip

\centerline{\bf Term 6 ($fff$)}

Using the first equation of (\ref{derB}) gives 
\bqy
6&:=:& \int \, f \left[ B\cdot\left(\p_v F_f\times\p_v G_f\right),H_f\right]_B\ncr
&:=:& \int \, f B\cdot\Big(\p_v\big(B\cdot\left(\p_v F_f\times\p_v G_f\right)\big)\times\p_v H_f 
\Big)\ncr
&:=:& \int \, f B_iB_r\ep_{ijk}\ep_{rst}\frac{\p}{\p v_j} \left(\frac{\p F_f}{\p v_s} \frac{\p G_f}{\p v_t}\right) 
\frac{\p H_f}{\p v_k} \ncr
&:=:& \int \, f B_iB_r\ep_{ijk}\ep_{rst}\left( 
 \frac{\p^2 F_f}{\p v_j\p v_s} \frac{\p G_f}{\p v_t} \frac{\p H_f}{\p v_k}
 +
  \frac{\p F_f}{\p v_s}   \frac{\p^2 G_f}{\p v_j\p v_t}   \frac{\p H_f}{\p v_k}
 \right)
 \label{pen}\\
 &:=:& \int \, f B_iB_r\ep_{ijk}\ep_{rst}\left( 
 \frac{\p^2 F_f}{\p v_j\p v_s} \frac{\p G_f}{\p v_t} \frac{\p H_f}{\p v_k}
- 
 \frac{\p H_f}{\p v_k}   \frac{\p^2 F_f}{\p v_s\p v_j}   \frac{\p G_f}{\p v_t}\,,
  \right):=:0
  \label{ult}
\eqy
where (\ref{ult}) follows from (\ref{pen}) by permuting the second term of  (\ref{pen}), shifting the indices according to $s\rightarrow k$, $k \rightarrow t$, $t\rightarrow j$,  $j\rightarrow s$, and $i \leftrightarrow r$, and using the antisymmetry of the Levi-Civita symbol.

Note,  the above procedure is common in this game and  of general utility, so it is recorded  in the following: 

\medskip

\noindent{\bf Lemma 1} {\it  If two terms can be made to cancel by permuting one of them, then all terms cancel. }

\proof   By writing out all six terms by permuting  
$FGH\rightarrow GHF\rightarrow HFG$, one observes they cancel in pairs.  \  $\Box$

\medskip

\noindent{\bf Remark.}  Term 6 vanishes without any assumptions on $B$, i.e.\ $\nabla\cdot B=0$ is \underline{not}  required.

 \medskip
 
 \centerline{\bf Term 11 ($EEf$)}
 
Using the first and last equations of (\ref{derEF})  yields
 \bqy
 11 &:=:&\int f \, H_E\cdot \frac{\p }{\p v} \frac{\de \Brac{F}{G}_{Ef}}{\de f}\ncr
 &:=:& \int f \, H_E\cdot  \p_v  \left(
G_E\cdot \p_v F_f - F_E\cdot \p_v G_f
\right)\ncr
&:=:& \int f \, \Big(
H_E\cdot  \p_v  \left(
G_E\cdot \p_v F_f \right)
- G_E\cdot  \p_v\left(H_E\cdot \p_v F_f
\right)
\Big):=: 0\,,
\eqy
where the last equality follows because   ${ com}(G_E\cdot\p_v,H_E\cdot\p_v)=0$, where  $com$ means commutator.

 \medskip
 
 \centerline{\bf Terms 2 and  5 ($fff$)}
 
\noindent{\bf Remark.}  Terms 2 and 5 have been grouped together because both give 
rise to terms in the Jacobi identity involving $F_f$, $G_f$, and $H_f$.  

\medskip

Using  (\ref{derC}) in $2 + 5 :=: \Brac{\Brac{F}{G}_c}H_B + \Brac{\Brac{F}{G}_B}H_c$ gives
\bqy
2 + 5 &:=:&\int f\, B\cdot
\left(
 {\p_v} \brac{F_f}{G_f}_c\times {\p_v} H_f
\right)
+
f \, \brac{B\cdot(\p_vF_f\times\p_vG_f)}{G_f}_c
\ncr
&:=:&\int \, f\, B_i\, \ep_{ijk} \frac{\p H_f}{\p v_k}
\left(
 \left[\frac{\p F_f}{\p v_j}, {G_f}\right]_c 
+ 
 \left[{F_f}, \frac{\p G_f}{\p v_j}\right]_c 
\right)\ncr
&+&\int \, f\,  \ep_{ijk}
\left(
\frac{\p}{\p x_{\ell}}
 \left(B_i\frac{\p F_f}{\p v_j} \frac{\p G_f}{\p v_k}
 \right)
 \frac{\p H_f}{\p v_{\ell}}
 -
 B_i
\frac{\p}{\p v_{\ell}}
 \left(
 \frac{\p F_f}{\p v_j} \frac{\p G_f}{\p v_k}
 \right) 
     \frac{\p H_f}{\p x_{\ell}}
     \right)\ncr
     &:=:&\int \, f\, B_i\, \ep_{ijk} 
\left(
\frac{\p H_f}{\p v_k}  \left[\frac{\p F_f}{\p v_j}, {G_f}\right]_c 
+ 
 \frac{\p H_f}{\p v_k} \left[{F_f}, \frac{\p G_f}{\p v_j}\right]_c 
 +   \left[\frac{\p F_f}{\p v_j} \frac{\p G_f}{\p v_k},  H_f \right]_c 
\right)
\label{null}\\
&+&\int \, f\,  \ep_{ijk} \frac{\p B_i}{\p x_{\ell}}
\left(
\frac{\p F_f}{\p v_j} \frac{\p G_f}{\p v_k} \frac{\p H_f}{\p v_{\ell}} 
\right)\,.
\label{resid}
\eqy

Upon defining $\calf=\p_vF_f$, $\calg=\p_vG_f$, and $\calh=\p_vH_f$, and using the Leibniz rule for $[\  ,\  ]_c$, Line (\ref{null}) can be rewritten as
\bq
\int \, f\, B_i\, \ep_{ijk} 
\big(
\calh_k \left[\calf_j, {G_f}\right]_c 
+ 
\calh_k\left[{F_f}, \calg_j\right]_c 
 +   \calg_k\left[\calf_j,  H_f \right]_c 
 +      \calf_j\left[\calg_k,  H_f \right]_c 
\big):=:0\,,
\eq
where upon permutation the first and fourth terms cancel, as do the second and third. Thus, 
\bq
2+5:=:\int \, f\,  \ep_{ijk} \frac{\p B_i}{\p x_{\ell}}
\left(
\frac{\p F_f}{\p v_j} \frac{\p G_f}{\p v_k} \frac{\p H_f}{\p v_{\ell}} 
\right)\,.
\label{near}
\eq
From (\ref{near}),  the only remaining term is from Line (\ref{resid}).  This term can be rearranged to yield
\bq
2 + 5 + {\rm cyc} =\int f\, \nabla\cdot B \left[ \left({\p_v F_f}\times {\p_v G_f}\right)\cdot {\p_v H_f}\right]\,,
\eq
which is a consequence of the following: 

\medskip

\noindent{\bf Lemma 2} {\it For any three vectors $\calf$, $\calg$, and $\calh$, and vector field $B$ in $\R^3$, 
\bq
\calh\cdot\nabla B\cdot (\calf\times\calg) + 
\calf\cdot\nabla B\cdot (\calg\times\calh) + 
\calg\cdot\nabla B\cdot (\calh\times\calf) 
= (\nabla\cdot B)\big(\calh\cdot(\calf\times\calg)\big)
\eq
where $\calh\cdot\nabla B\cdot \calc=\calh_i \calc_j \p_iB_j$.}

\medskip
\noindent{\bf Remark.}  Terms 2 and 5  could have been combined with Term 6.  They were  considered  separately to pinpoint, as  will be seen,  that they are the sole terms that  violate the Jacobi identity without $\nabla \cdot B=0$.

 \medskip
 
 \centerline{\bf Terms 7 and  10 ($Eff$)}

Inserting the first and last equations of (\ref{derB}) and the first equation of (\ref{derEF}) into 
$7 + 10 :=: \Brac{\Brac{F}{G}_B}H_{Ef} + \Brac{\Brac{F}{G}_{Ef}}H_B$ gives
 \bqy
7 + 10 &:=:& \int \, f\, H_E\cdot\p_v 
\big(
B\cdot 
\left(
\p_v F_f \times \p_v G_f
\right)
\big) \ncr
&{\ }&\hspace{1 cm} +  f \, B\cdot
\big(
\p_v \left(
-G_E\cdot\p_v F_f + F_E\cdot \p_v G_f
\right)
\times \p_v H_f )
\big)
\ncr
&:=:& 
\int\,  f\,
{H_{E}}_{\ell}{\p_v}_{\ell}
\left(
B_i \ep_{ijk}{\p_v}_j F_f {\p_v}_k G_f
\right)
\ncr
&{\ }&\hspace{1 cm} - f\,  B_i \, \ep_{ijk}
\left(
- {\p_v}_j ({G_{E}}_{\ell}{\p_v}_{\ell} F_f){\p_v}_k H_f
+ {\p_v}_j ({F_{E}}_{\ell}{\p_v}_{\ell} G_f){\p_v}_k H_f
\right)
\ncr
&:=:& 
\int\,  f\,
B_i \ep_{ijk}{H_{E}}_{\ell}
\left(
 {\p_v}_k G_f {\p_v}_{\ell}{\p_v}_j F_f+ {\p_v}_j F_f {\p_v}_{\ell} {\p_v}_k G_f
\right)
\label{7+10b}\\
&{\ }&\hspace{1 cm} - f\,  B_i \, \ep_{ijk}
\left(
-  {G_{E}}_{\ell}{\p_v}_k H_f {\p_v}_j {\p_v}_{\ell} F_f
+ {F_{E}}_{\ell}  {\p_v}_k H_f{\p_v}_j {\p_v}_{\ell} G_f
\right)\,.
\label{7+10c}
\eqy
 Now upon permutation,  the first term of (\ref{7+10b}) is seen to cancel the second of (\ref{7+10c}) and the second term of (\ref{7+10b}) cancels the first term of (\ref{7+10c}).  Thus $7 + 10 :=: 0$.
 
\medskip

\centerline{\bf Terms 3, 8,  and  9 ($Eff$)}

Using the first and last equations of (\ref{derC}) and (\ref{derEF}) and the first and second equations of (\ref{derB}) 
in 3 + 9 gives 
\bq
3+9 :=:-\int f \left(
H_E\cdot\p_v\left[F_f,G_f\right]_c + 
\left[H_E\cdot\p_vG_f,F_f\right]_c -
\left[H_E\cdot\p_vF_f,G_f\right]_c 
\right)
\label{pen39}
\eq
Using Lemma 3 below  in (\ref{pen39}) with  $C$ equal to  $H_E$ gives
\bqy
3+9 &:=:&\int f \, (\nabla\times H_E)\cdot \left(\p_v F_f\times\p_v G_f\right)\,.
\label{3+9}
\eqy

\medskip

 \noindent{\bf Lemma 3} {\it For any vector field $C(x)$ and phase  space functions $f$ and $g$, 
\bq
 C\cdot\p_v\brac fg_c=\brac{C\cdot\p_v f}{g}_c + \brac{f}{C\cdot\p_v g}_c 
 +  (\nabla \times C)\cdot\left(\p_vg\times\p_vf\right)\,.
\eq
}
\proof  
With ${\p_{v}}_i:= \p/\p v_i$ and ${\p_{x}}_i:= \p/\p x_i$, 
\bqy 
C\cdot\p_v\brac fg_c&=& C_i\big(
\left[ {\p_v}_i f, g\right]_c + \left[f, {\p_v}_ig\right]_c
\big)
\ncr
&=&\left[C\cdot\p_v f,g\right]_c -  \left[C_i, g\right]_c{\p_v}_i f
+ \left[f,C\cdot\p_v g\right]_c -  \left[f,C_i\right]_c{\p_v}_i g
\ncr
&=&\left[C\cdot\p_v f,g\right]_c -  
({\p_v}_i f)({\p_v}_j g){\p_x}_j C_i
+ \left[f,C\cdot\p_v g\right]_c +({\p_v}_i g)({\p_v}_j f){\p_x}_j C_i \ncr
&=&\left[C\cdot\p_v f,g\right]_c
+ \left[f,C\cdot\p_v g\right]_c 
+({\p_x}_j C_i)\big(
({\p_v}_i f)({\p_v}_j g)-
({\p_v}_i g)({\p_v}_j f)
\big)\ncr
&=&\left[C\cdot\p_v f,g\right]_c
+ \left[f,C\cdot\p_v g\right]_c 
+({\p_v}_i g)({\p_v}_j f)
\big(
{\p_x}_j C_i - {\p_x}_iC_j
\big)\ncr
&=&\brac{C\cdot\p_v f}{g}_c + \brac{f}{C\cdot\p_v g}_c 
 +  (\nabla \times C)\cdot\left(\p_vg\times\p_vf\right)\,. \ \Box
\eqy

Now consider Term 8
 \bqy
 8 &:=:&-\int\! d^3x \, H_E\cdot \nabla\times\left(\int\! d^3v \, f\, \p_v F_f\times\p_v G_f\right)\ncr
 &:=:&-\int\! d^6z \, f\, (\nabla\times H_E)\cdot  (\p_v F_f \times \p_v G_f )\,.
 \label{8}
 \eqy 
 Equations (\ref{3+9}) and (\ref{8}) imply   $3 + 8 + 9 + {\rm cyc} =0$. 
 
 \medskip
 
\noindent{\bf Remark.}  Observe the terms here, like Terms 7 and 10,  are $Eff$ terms. However, they have been grouped separately because  there are `other considerations' as mentioned above.  The Terms 7 and 10 vanish with $B$, but the Terms 3, 8, and 9, do not.  Thus terms of one kind cannot cancel terms of the other. 

\medskip

Finally, from all of the above,  the following  is concluded:

\medskip

\noindent{\bf Main Theorem}  (\cite{morrison82}) {\it For the Vlasov-Maxwell bracket of (\ref{MVbktS})
\bq
\nest FGH + {\rm cyc}= \int\!d^6z \, f\, 
\nabla\cdot B \,  \left({\p_v F_f} \times {\p_v G_f}\right)\cdot {\p_v H_f} \,.
\label{main}
\eq } 

\medskip

\noindent{\bf Remark.}  It is interesting to note that the other constraint, $\nabla\cdot E=4\pi \rho$,  need not be satisfied for the Jacobi identity to hold.  It turns out to be a Casimir invariant. 
 

\subsection{Jacobi identity for the spin Vlasov-Maxwell bracket}
\label{SpinJacobi}

Writing $\{F,G\}_{sVM}=\{F,G\}_{VM} + \{F,G\}_{s}$ and using $:=:$ as defined in Appendix \ref{MVjacobi}
\bqy
\{\{F,G\}_{sVM},H\}_{sVM}&:=:& \{\{F,G\}_{VM},H\}_{VM} + \{\{F,G\}_{s},H\}_{VM} +
\nonumber\\
&{\ }& \qquad+ \{\{F,G\}_{VM},H\}_{s} + \{\{F,G\}_{s},H\}_{s}
\nonumber\\
&:=:& \{\{F,G\}_{s},H\}_{VM} 
+ \{\{F,G\}_{VM},H\}_{s}\,,
\eqy
where the second equality follows because of the Jacobi identity for Vlasov-Maxwell (assuming solenoidal $B$) and the fact that $\{F,G\}_{s}$ is a Lie-Poisson bracket (see e.g.\ \cite{morrison98,marsden}).  Thus it only remains to show that the cross terms cancel, which is facilitated by a the bracket theorem \cite{morrison82} stated in Appendix \ref{Ajacobi}; viz., when functionally differentiating $\{F,G\}_{VM}$ and $\{F,G\}_{s}$, which are needed when constructing the cross terms,  one can ignore the second functional derivative terms.  Using the symbol $\deq$ again to denote equivalence modulo the second variation terms, 
\bqy
\frac{\de \{F,G\}_{VM}}{\de f}&\deq& [F_f,G_f]_c + [F_f,G_f]_B
+ F_E\cdot\p_v G_f - G_{E}\cdot\p_v F_f 
\\
\frac{\de \{F,G\}_{s}}{\de f}&\deq& [F_f,G_f]_s  
\eqy
while all other needed functional derivatives vanish.  Thus
\bqy
\{\{F,G\}_{VM},H\}_{s}&:=:& \int\!\!d^9z\,\Big(f \, \big[[F_f,G_f]_c+ [F_f,G_f]_B, H_f\big]_s
\nonumber\\
&{\ }& \hspace{1.0 in} +   f \big[F_E\cdot\p_v G_f - G_{E}\cdot\p_v F_f,H_f\big]_s\Big)
\label{jac1}\\
\{\{F,G\}_{s},H\}_{VM}&:=:& \int\!\!d^9z\,  \Big( f \big[[F_f,G_f]_s, H_f\big]_c +  f \big[[F_f,G_f]_s, H_f\big]_B\Big)
\nonumber\\
&{\ }& \hspace{1.0 in} +    f \,  H_E\cdot\p_v [F_f,G_f]_s\Big)
\label{jac2}
\eqy
The first lines of (\ref{jac1}) and (\ref{jac2}) cancel by virtue of the Jacobi identities for the brackets $[\, ,\, ]_{c,B,s}$ on functions, while the second line of (\ref{jac1}) cancels upon permutation of  the second term. Similarly, the second term of (\ref{jac2}) vanishes. 


\subsection{Jacobi identity for the monopole Vlasov-Maxwell bracket}
\label{MonopoleJacobi}

For this case the Gaussian units of the text are used, i.e.\   the factors of $4\pi$ are reinserted and  both the usual and monople charges are manifest.  Let  $\Brac FG_m=\Brac{F}{G}_{VM} + \Brac{F}{G}_{M}$,  where
\bq
\Brac{F}{G}_M= \int \! d^6z \, f\brac{F_f}{G_f}_E +  \frac{4\pi g}{m}\, f \left(G_B\cdot\p_vF_f - F_B\cdot\p_v G_f\right) \,,
\label{mbkt}
\eq
and $\brac{\ \,   }{\ }_E$ is defined by (\ref{Ebkt}).  Thus, the Jacobi identity has four terms to consider
\bqy
\Brac{\Brac{F}{G}_m}{H}_m&:=:&\Brac{\Brac{F}{G}_{VM}}{H}_m + \Brac{\Brac{F}{G}_M}{H}_m  
 \nonumber\\
  &=&
  \underset{1}{\underbrace{\Brac{\Brac{F}{G}_{VM}}H_{VM}}}
 + \underset{2}{\underbrace{\Brac{\Brac{F}{G}_{M}}H_{VM}}}\ncr
 &+& \underset{3}{\underbrace{\Brac{\Brac{F}{G}_{VM}}H_{M}}}
  +\underset{4}{\underbrace{\Brac{\Brac{F}{G}_{M}}H_{M}}}\,,
  \label{jack}
  \eqy
where the symbol $:=:$ is defined in Appendix \ref{MVjacobi}. 
 
\medskip

\centerline{\bf Term 1}

From the \cite{morrison82} (cf.\ Appendix \ref{MVjacobi})
\bq
\{\{F,G\}_{VM},H\}_{VM} + {\rm cyc}= \frac{ e}{m^2}\int\!d^6z \, f\, 
\nabla\cdot B \left[ \left({\p_v F_f}\times {\p_v G_f}\right)\cdot  {\p_v H_f}\right]\,.
\label{old}
\eq

As in Appendix \ref{MVjacobi},    $\de\Brac{F}{G}/\de \psi \deq \, \dots$ denotes the functional derivative modulo the second derivative terms. The following will be needed:
\bqy
\frac{\de \Brac{F}{G}_{VM}}{\de f}&\deq&
\brac{F_f}{G_f}_c +  \brac{F_f}{G_f}_B 
+\frac{4\pi e}{m}\, \left(G_E\cdot\p_vF_f - F_E\cdot\p_v G_f\right)
\\
\frac{\de \Brac{F}{G}_{M}}{\de f}&\deq&
   \brac{F_f}{G_f}_E 
+\frac{4\pi g}{m}\,  \left(G_B\cdot\p_vF_f - F_B\cdot\p_v G_f\right)
\label{Mf}
\\
\frac{\de \Brac{F}{G}_{VM}}{\de B}&\deq& \frac{e}{m^2}\int \! d^3v\,  f \, {\p_v  F_{f} } \times
 {\p_v  G_{f}}  \,, \quad 
\quad
\frac{\de \Brac{F}{G}_{VM}}{\de E}\deq 0
\\
\frac{\de \Brac{F}{G}_{M}}{\de E}&\deq& -\frac{g}{m^2} \int \! d^3v\,  f \, {\p_v F_{f}}  \times
{\p_v G_{f}  }  \,, \quad 
\quad
\frac{\de \Brac{F}{G}_{M}}{\de B}\deq 0\,.
\eqy

The Poisson bracket that generates  the `$v\times$' part of the generalized  Lorentz force is
$\brac{f}{g}=\brac{f}{g}_c + \brac{f}{g}_B  + \brac{f}{g}_E$.  Because  of 
\bq
\brac{\brac{f}{g}_E}{h}_E + {\rm cyc}=0
\quad {\rm and}\quad 
\brac{\brac{f}{g}_E}{h}_B + \brac{\brac{f}{g}_B}{h}_E + {\rm cyc} =0\,.
\label{EB}
\eq
the following holds: 
\bqy
\brac{\brac{f}{g}}{h} + {\rm cyc}
&=& 
\brac{\brac{f}{g}_c}{h}_B + \brac{\brac{f}{g}_B}{h}_c
+
\brac{\brac{f}{g}_c}{h}_E + \brac{\brac{f}{g}_E}{h}_c
+ {\rm cyc}\ncr
&=& \left({e} \nabla \cdot B - {g} \nabla \cdot E\right)
\left[ \left({\p_v f} \times {\p_V g}\right) \cdot {\p_V h}\right]/m^2
\label{bad}
\eqy
The first term of the above is the source of the RHS  of (\ref{old}). 

Now  consider   the remaining terms of (\ref{jack}). 

\medskip

\centerline{\bf Term 4}

Equations (\ref{Mf}) and (\ref{mbkt}) give 
\bqy
\{\{F,G\}_M,H\}_M&:=:&\int d^6z\, f\,\left[
\brac{F_f}{G_f}_E 
+\frac{4\pi g}{m}\,  \left(G_B\cdot\p_vF_f - F_B\cdot\p_v G_f\right),H_f
\right]_E\ncr
&{\ }& \hspace{-.5 in} +\,  \frac{4\pi g}{m} f \,H_B\cdot\p_v\left(
\brac{F_f}{G_f}_E 
+\frac{4\pi g}{m}\,   \left(G_B\cdot\p_vF_f - F_B\cdot\p_v G_f\right)
\right)\,.
\label{MMJ}
 \eqy

Equation (\ref{MMJ}) has three kinds of terms. Consider first the $fff$-terms:
\bqy
 \int d^6z\, f\,\left[
\brac{F_f}{G_f}_E ,H_f
\right]_E&=&
\frac{g^2}{m^4}
 \int d^6z\, f\,E\cdot
 \Big(\p_v
 \big(
E\cdot\left( \p_v F_f\times \p_v G_f
\right)\big)\times \p_v H_f
\Big)
\ncr
&=&
\frac{g^2}{m^4}
 \int d^6z\, f\,\ep_{ijk}\ep_{rst} E_iE_r{\p_v}_kH_f
 \left(
 {\p_v}_j{ \p_v}_s F_f { \p_v}_tG_f \right.\ncr
 &{\ }&\hspace{1.0 in} \left.+ {\p_v}_j{ \p_v}_tG_f { \p_v}_sF_f
 \right)
 \label{4}
\eqy
Upon permutation and reindexing,  the two terms of (\ref{4}) cancel.  This, of course, follows 
immediately from (\ref{EB}) -- the above serves as a proof that $\brac{\brac{f}{g}_E}{h}_E + {\rm cyc}=0$. 
Next consider the $BBf$-term:
\bq
\frac{16\pi^2g^2}{m^4}
 \int d^6z\, f
 H_B\cdot\p_v
 \left(G_B\cdot\p_vF_f - F_B\cdot\p_v G_f\right)
\eq
This vanishes upon permutation because $H_B\cdot\p_v$ and $F_B\cdot\p_v$ commute. 
Now all that remains of Term 4 is the $Bff$-term:
\bq
\frac{4\pi g}{m} \int d^6z\, f\,\left[
 (G_B\cdot\p_vF_f - F_B\cdot\p_v G_f ,H_f
\right]_E +   f \,H_B\cdot\p_v 
\brac{F_f}{G_f}_E 
\,.
 \eq
This term is of the same form  as the $EFF$ term of the MV-bracket (terms 7 and 10),
 and vanishes for the same reason.  Therefore, Term 4 vanishes.

Now consider the two cross terms.

\medskip

\centerline{\bf Terms 2 and 3}

Term 2 is
\bqy
\{\{F,G\}_M,H\}_{VM}&:=:&\int d^6z\, f\,\left[
\brac{F_f}{G_f}_E 
+\frac{4\pi g}{m}\,  \left(G_B\cdot\p_vF_f - F_B\cdot\p_v G_f\right),H_f
\right]_c
\label{mmv1}\\
&+&  f\,\left[
\brac{F_f}{G_f}_E 
+\frac{4\pi g}{m}\,  \left(G_B\cdot\p_vF_f - F_B\cdot\p_v G_f\right),H_f
\right]_B
\label{mmv2}\\
&+& \frac{4\pi e}{m} f \,H_E\cdot\p_v\left(
\brac{F_f}{G_f}_E 
+\frac{4\pi g}{m}\,   \left(G_B\cdot\p_vF_f - F_B\cdot\p_v G_f\right)
\right)
\label{mmv3}\\
&+& \frac{e}{m} f\, \p_v H_f \cdot \frac{g}{m^2} \int \! d^3 v\,    f \, {\p_v F_{f}} \times
{\p_v G_{f}}
\label{mmv4}\\
&-& 4\pi \int d^3x \, \nabla \times H_B \cdot 
\frac{g}{m^2} \int \! d^3  v\,    f \,  {\p_v F_{f}}  \times
 {\p_v G_{f}}\,,
\label{mmv5}
 \eqy
while Term 3 is
\bqy
\{\{F,G\}_{VM},H\}_{M}
&:=:&\int d^6z\, f\,\Big[
\brac{F_f}{G_f}_c +  \brac{F_f}{G_f}_B  
\label{mvm1}\\
&{\ }& \hspace{ 1 cm} + \frac{4\pi e}{m}\, \left(G_E\cdot\p_vF_f - F_E\cdot\p_v G_f\right),H_f
\Big]_E
\label{mvm2}\\
&+& \frac{4\pi g}{m} f \,H_B\cdot\p_v\Big(
\brac{F_f}{G_f}_c +  \brac{F_f}{G_f}_B 
\label{mvm3}\\
&{\ }& \hspace{ 1 cm} + \frac{4\pi e}{m}\, \left(G_E\cdot\p_vF_f - F_E\cdot\p_v G_f\right)\Big)
\label{mvm4}\\
&-& \frac{g}{m} f\, \p_v H_f \cdot \frac{e}{m^2} \int \! d^3 v\,   f \,  {\p_v F_{f}}  \times
 {\p_v G_{f}} \,.
\label{mvm5}
 \eqy

Upon comparing Terms 2 and 3 some cancellations are immediate.
\begin{itemize}
\item Using (\ref{bad}), 1st term of (\ref{mmv1}) + 1st term of (\ref{mvm1}) gives 
$
- \frac{g}{m^2}( \nabla \cdot E )
  \left(\p_v F_f\times \p_v G_f\right)\cdot \p_v H_f
$
\item   Using (\ref{EB}), 1st term of (\ref{mmv2}) + 2nd term of (\ref{mvm1}) = 0 
\item  Lines (\ref{mmv4}) +  (\ref{mvm5}) =0
\item The terms of (\ref{mvm4}) vanish because $H_B\cdot \p_v$ and  $G_E\cdot \p_v$ commute.  Likewise the last two terms of (\ref{mmv3}) 
\end{itemize}

Applying the following:

\medskip 
 
 \noindent{\bf Lemma} {\it For any vector field $C(x)$ and phase  space functions $f$ and $g$, 
\bq
 C\cdot\p_v\brac fg_c=\brac{C\cdot\p_v f}{g}_c + \brac{f}{C\cdot\p_v g}_c 
 + m^{-1} (\nabla \times C)\cdot\left(\p_vg\times\p_vf\right)\,,
\eq
}
which is not difficult to prove, to the last terms of  (\ref{mmv1}) + Line (\ref{mmv5})   
+ the last first term of (\ref{mvm3}) =0.

There are six remaining terms.  The last two terms of (\ref{mmv2}) cancel the last term of (\ref{mvm3}), and the first   term of (\ref{mmv3}) cancels the  two terms of (\ref{mvm2}).   Thus  in this case,  the obstruction becomes 
\bqy
\{\{F,G\}_{mVM},H\}_{mVM} + {\rm cyc}&=&\\
&{\ }& \hspace{-.75 cm}  \frac1{m^2}\int\!d^6z \, f\, 
\left(
{e}\nabla\cdot B-  {g} \nabla\cdot E
\right)
\left(\p_{v} F_f\times \p_{v} G_f \right)\cdot \p_{v} H_f\,, 
\nonumber
\eqy
and it is concluded that  the Jacobi  identity still requires a solenoid constraint on $e B-  {g} E$. 

 Upon transforming to new variables $\tilde{e} \tilde{E}= eE + gB$ and $\tilde{e} B= eB-gE$, where $\tilde{e}^2=e^2 + g^2$,   reproduces the Poisson bracket for Vlasov-Maxwell theory, which is possible  for this single species case of Dirac's theory.  In retrospect, the existence of this transformation precludes the necessity for the proof of the Jacobi identity; however, consistent with the goal of this entire  appendix, viz.\  to demonstrate  techniques  of general utility rather than to present the most efficient proofs, we retain it here.


\section*{Acknowledgment}
\noindent  I would like to thank Alain Brizard, Cristel Chandre, Emanuele Tassi, and Michel Vittot  for their continued  interest in this subject, their encouragement, and for many fruitful discussions. I would also like to  thank  Iwo Bialynicki-Birula for  helpful  correspondence and Loic de Guillebon for his comments on an earlier draft of this manuscript.  Supported by U.S. Dept.\ of Energy Contract \# DE-FG05-80ET-53088.

\bibliographystyle{apsrev}


\end{document}